\documentclass[journal,comsoc]{IEEEtran}
\usepackage[table]{xcolor}

\usepackage[T1]{fontenc}
\usepackage{verbatim}
\ifCLASSINFOpdf
\else
\fi

\usepackage{amsmath}
\usepackage{bm}
\usepackage[cmintegrals]{newtxmath}

\usepackage{balance}
\usepackage{mwe}
\usepackage[caption=false]{subfig}
\usepackage{float}
\usepackage{url}
\usepackage{tabularx}
\usepackage{tikz}
\usepackage[hidelinks]{hyperref}
\usepackage{adjustbox}
\usepackage{booktabs}
\usepackage{rotating}
\usepackage{slashbox}
\usepackage{xcolor}
\usepackage{color}
\usepackage{colortbl}
\usepackage{multirow}
\usepackage[misc,geometry]{ifsym}
\usepackage{dirtytalk}
\usepackage[normalem]{ulem}
\usepackage{color}

\usepackage{svg}
\usepackage{amsmath} 
\usepackage{physics} 
\usepackage{blindtext}
\usepackage[ruled,vlined]{algorithm2e}
\newcommand\Tstrut{\rule{0pt}{2.6ex}}       
\newcommand\Bstrut{\rule[-1.5ex]{0pt}{0pt}} 
\newcommand{\TBstrut}{\Tstrut\Bstrut} 
\definecolor{grey}{RGB}{197,197,197}
\newcommand{\orcid}[1]{\href{https://orcid.org/#1}{\includesvg[width=10pt]{orcid}}}
\usepackage{color, colortbl}
\definecolor{grey}{gray}{0.9}
\usepackage{cite}
\usepackage{lineno}
\usepackage[bottom]{footmisc}
\usepackage{graphicx}
\usepackage{indentfirst}
\usepackage{makecell}
\colorlet{mygreen}{green!60!gray}

\usepackage{tablefootnote}
\usepackage{threeparttable}
\usepackage{background}
\backgroundsetup{contents=Published in IEEE TNSM, opacity=0.45, scale=8, color=gray, angle=60}

\begin{document}

\title{Employing Deep Ensemble Learning for Improving
	the Security of Computer Networks against
	Adversarial Attacks}

\author{
Ehsan~Nowroozi,~\IEEEmembership{Senior Member,~IEEE,} Mohammadreza Mohammadi,~\IEEEmembership{Member,~IEEE,}\\Erkay Savaş,~\IEEEmembership{Member,~IEEE,} Mauro Conti,~\IEEEmembership{Fellow Member,~IEEE,} and Yassine Mekdad,~\IEEEmembership{Member,~IEEE}
\IEEEcompsocitemizethanks{
\IEEEcompsocthanksitem E. Nowroozi is with Bahcesehir University, Faculty of Engineering and Natural Sciences, Computer Engineering Department, Istanbul, Turkey (Email: ehsan.nowroozi65@gmail.com, ehsan.nowroozi@eng.bau.edu.tr).
\IEEEcompsocthanksitem M. Mohammadi and M. Conti are with the Department of Mathematics, Security \& Privacy Research Group (SPRITZ), University of Padua, 35121, Padua, Italy. (Email: {mohammadreza.mohammadi@studenti.unipd.it}, {mauro.conti@unipd.it})
\IEEEcompsocthanksitem E. Savaş, is a Faculty of Engineering and Natural Sciences (FENS), Sabanci University, Istanbul Turkey 34956 (Email: ehsan.nowroozi@sabanciuniv.edu, erkays@sabanciuniv.edu)
\IEEEcompsocthanksitem Y. Mekdad is with the Cyber-Physical Systems Security Lab, Department of Electrical and Computer Engineering, Florida International University, Miami, FL, 33174. (Email: ymekdad@fiu.edu)
}
}
\maketitle

\begin{abstract}
In the past few years, Convolutional Neural Networks (CNN) have demonstrated promising performance in various real-world cybersecurity applications, such as network and multimedia security. However, the underlying fragility of CNN structures poses major security problems, making them inappropriate for use in security-oriented applications including such computer networks. 
Protecting these architectures from adversarial attacks necessitates using security-wise architectures that are challenging to attack.

In this study, we present a novel architecture based on an ensemble classifier that combines the enhanced security of 1-Class classification (known as 1C) with the high performance of conventional 2-Class classification (known as 2C) in the absence of attacks.
Our architecture is referred to as the 1.5-Class (SPRITZ-1.5C) classifier and constructed using a final dense classifier, one 2C classifier (i.e., CNNs), and two parallel 1C classifiers (i.e., auto-encoders). In our experiments, we evaluated the robustness of our proposed architecture by considering eight possible adversarial attacks in various scenarios. We performed these attacks on the 2C and SPRITZ-1.5C architectures separately. The experimental results of our study showed that the Attack Success Rate (ASR) of the I-FGSM attack against a 2C classifier trained with the N-BaIoT dataset is 0.9900. In contrast, the ASR is 0.0000 for the SPRITZ-1.5C classifier. 
\end{abstract}

\section{Introduction}

\IEEEPARstart{T}{}he development of reliable Machine Learning (ML) techniques with a particular reference to Computer Networks applications is gaining popularity. Such technology is capable of providing satisfactory performance in the face of an adversary trying to impede an accurate analysis. In this context, Digital Forensics (DF), which is a field of study in academia, tries to gather data on the history of documents, their authenticity, source, the process they underwent, and other factors. However, complicated forensics tasks often need statistical features and modeling. Therefore, forensic scientists have to consider different ML techniques. Whereas disabling ML/DL-based forensic analysis turns out to be a simple task. In this case, the adversary is trying to employ different Counter-Forensics (CF) techniques to erase the main traces that the ML detector relies on for distinguishing, such as pristine and malicious examples~\cite{Gloe2007, bohme2013counter, nowroozi2020machine, NOWROOZI2021102092, Ehsan_URL_Detection}.   
In recent years, various CF approaches have been developed to bypass forensics analysis and are often commonly referred to as adversarial attacks. As a consequence, the primary objective of an adversary is to impede detection by removing or altering evidence of illicit processing and making the counterfeited example look authentic \cite{nowroozi2020machine, NOWROOZI2021102092}. 
This is particularly the case for ML-Based approaches. As explained in \cite{hernandez2022adversarial}, and highlighted in \cite{NOWROOZI2021102092} in the context of Adversarial Machine Learning (adv-Ml), in reality, the adversary may utilize the fragility of ML techniques and lunch different attack methodologies. 

The information collected by the adversary concerning the machine to be attacked will have a significant impact on the adversary's strategies and the countermeasures employed by the forensic specialist to counter them \cite{NOWROOZI2021102092}. 
A Perfect Knowledge (PK) setting, according to [7], is a situation in which the adversary has full knowledge of the forensic approach being attacked. This is the complete information from the training phase of the ML model in the domain of ML-based methodologies. In a PK scenario, powerful CF attacks can be used to impede correct analysis while causing a minor distortion in the attack example. In a PK setting, developing effective anti-CF methods is challenging since the attacker always makes the last move \cite{Ehsan_Eusipco_2017, Ehsan_IWBF}.
In practice, PK implies that the adversary knows any plausible defenses the analysts can perform. In this situation, the only possibility is to employ secure architecture methodologies, which are inherently more resilient against adversarial attacks. 

In this study, we explore Computer Networks detection by considering different DL models, which is a forensic analysis used to identify if an example has been subject to any manipulation. Therefore, the analysis is often known as a binary decision, meaning that the examples are divided into two areas, referring to pristine and malicious examples.
The most common approach in ML regarding this issue is to train the detector with malicious and pristine examples using many examples. 
The objective of the training procedure is to divide the training examples so that they may be appropriately classified. Furthermore, the detector should generalize well when met with new instances that were not observed during the training process. 

\begin{figure*}[ht]%
\centering
	\subfloat[2C]{\label{figRegion:a}\includegraphics[width=0.25\textwidth]{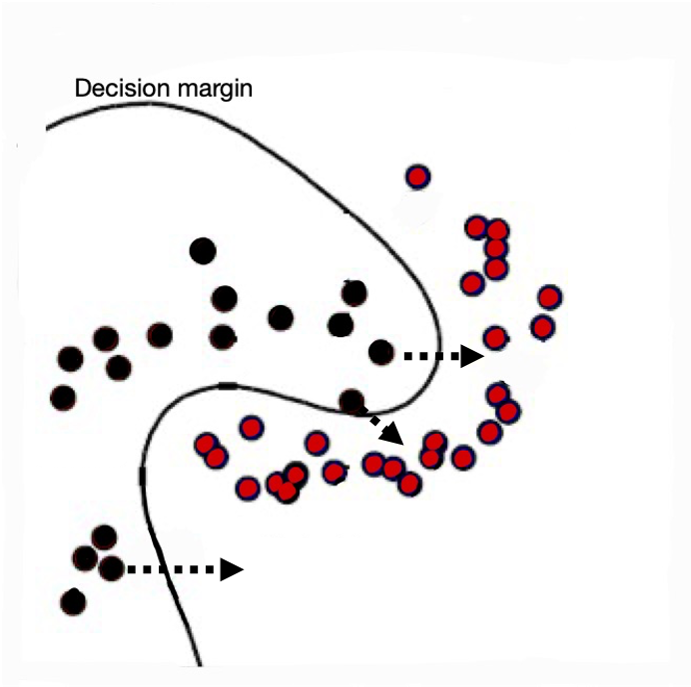}}
	\subfloat[1C]{{\label{figRegion:b}\includegraphics[width=0.25\textwidth]{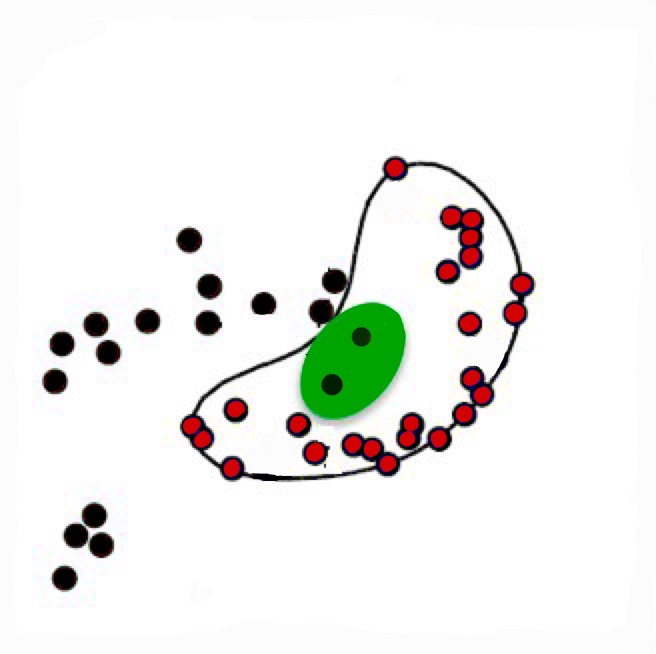} }}
	\subfloat[SPRITZ-1.5C]{{\label{figRegion:c}\includegraphics[width=0.25\textwidth]{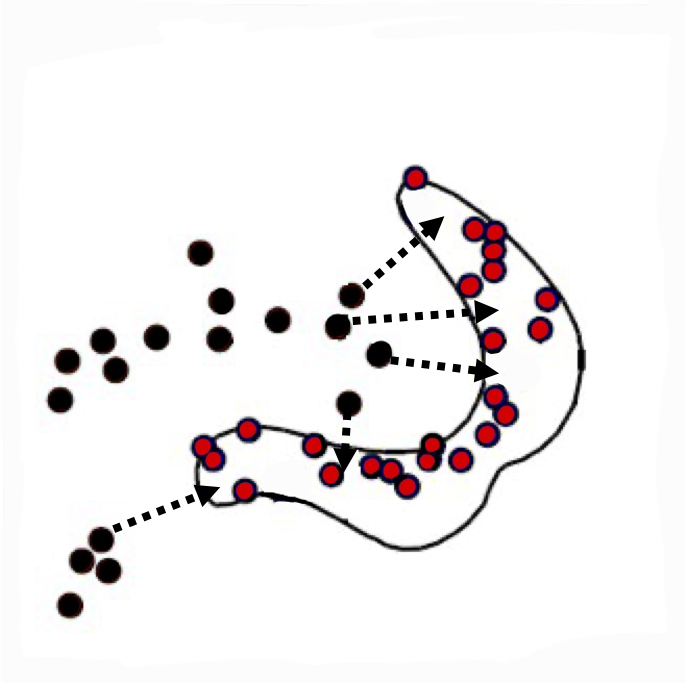} }}
	\caption{The 2C, 1C, and SPRITZ-1.5C classifiers' decision margins are visually presented. Red samples are considered pristine, whereas black dots signify malicious examples. The attacker attempts to relocate the black samples from the malicious region within the red region, a) The attacker takes full advantage of the availability of empty spaces to cause a missed detection error; as a result, the 2C classifier is vulnerable, b) Since they define a closed region enclosing samples from only one class, 1C classifiers are inherently more resilient to attacks with minor classification errors (a green circle in the figure highlights the examples that have a missed detection error), c) the SPRITZ-1.5C classifier is designed to exhibit similar resilience against attacks as the 1C classifier, but the acceptance region is better formed than in the 1C classification scenario.}%
	\label{figRegion:setup}%
\end{figure*}

In the procedure described before, the drawback is that the analyzer must choose between two classes of instances even when encountered with examples that differ significantly from those presented during the training phase, potentially because such examples do not belong to either. 
As soon as the examples fit into any of the aforementioned classifications, the classifier will return a true response depending on the instances seen during training. 
Given that these characteristics were not present during the training phase, the classifier would likely choose a random decision. This is unlikely to be an issue under normal work conditions because the classifier model will never require to handle such abnormal conditions. However, under adversarial settings, the adversary might use the availability of unpopulated areas of the example space to induce a low-cost classification error (we refer to it as distortion). 
Figure \ref{figRegion:a} depicts the above scenario, in which black example points are displaced with minimum distortion into an unpopulated red region.
The problem indicated above may be addressed using a 1C classifier when the adversary's goal is unidirectional~\cite{kang2022using}.
In this case, only examples from the specific class are employed to train the ML to evaluate whether an example belongs to that class. 
The significant distinction between a conventional 2C and a 1C classifier is that the latter divides the example region into a closed region with examples from the class considered during training and a complement region including all additional examples. As depicted in Figure~\ref{figRegion:b}, an attacker intending to move an example from the outside (black) region into the closed area containing the target class's examples (red region) can no longer take full advantage of unpopulated portions of the example area.

Considering the 2C superior accuracy advantage and 1C classifiers in robustness, as illustrated in Figures~\ref{figRegion:a} and ~\ref{figRegion:b}, it is required to construct a system that leverages the accuracy and robustness from 2C and 1C classifiers. As a result, the SPRITZ-1.5C classifier's main objective is to maintain both benefits: 
(i) accuracy of detection, and (ii) robustness, as shown in Figure~\ref{figRegion:c}. In this study, we propose the SPRITZ-1.5C classifier that has a comparable robustness against numerous adversarial attacks as the 1C classifier has; however, the acceptance regions are significantly formed in comparison to the 1C (see Figure~\ref{figRegion:c}). Note that the accuracy in the absence of attacks is similar to 2C classifier. This characteristic is proven in our study by considering two well-respected datasets: N-BaIoT and RIPE datasets. Despite its security appeal, the 1C classifier has limitations. These constraints of 1C may not employ any information about the examples relating to a particular two classes. A 1C may be developed by evaluating only pristine examples (i.e., 1C-Leg), but this ignores details about the specific traces left in a malicious example. In contrast, a 1C classifier may be developed by evaluating only malicious examples (i.e., 1C-Mal). Consequently, in the absence of attacks, the efficiency of 1C classifiers is expected to be lower than that of a 2C architecture; however, when compared to the approach in \cite{biggio2013evasion, barni2020improving}, we achieved good performance by including DL architectures. The authors in \cite{biggio2013evasion, barni2020improving} proposed a multi-classifier model based on a Support Vector Machine (SVM) that includes the benefits of both models, combining the greater accuracy of 2C classifiers with the inherent security of 1C classifiers. We considered this scenario and introduced SPRITZ-1.5C, a novel architecture based on a combination of multiple DL architectures (e.g., CNN as 2C classifier, pristine auto-encoder as one 1C-Leg, malicious auto-encoder as one 1C-Mal, and dense classifier). 




\subsection{Contributions}
Our contributions are outlined in the following:
\begin{itemize}
    \item 
   We propose an ensemble architecture SPRITZ-1.5C by employing deep ensemble learning in Computer Networks. The SPRITZ-1.5C architecture consists of four classifiers. Namely: 2C classifier (i.e., CNN), 1C classifier (i.e., trained with pristine examples), 1C classifier (i.e., trained with malicious examples), followed by a final dense classifier.
    
    \item 
    We perform eight adversarial attacks with different parameters against 2C (i.e., scenario 1) and SPRITZ-1.5C (i.e., scenario 2) classifiers separately: the I-FGSM, FGSM, JSMA, L-BFGS, PGD, BIM, DeepFool, and C\&W attack. We considered two well-known datasets in Computer Networks while performing these attacks: the RIPE-Atlas\cite{analysis} and N-BaIoT\cite{nbaiot} datasets.  
    

    \item 
    
    The impact of the selected adversarial attacks on the resilience of the 2C classifier and SPRITZ-1.5C architecture is systematically investigated. These attacks are considered during the testing phase, and we refer to them as exploratory evasion attacks.

    \item 
    
    
    We compared the security of the SPRITZ-1.5C architecture with 2C. To elaborate more, we demonstrate that the ASR for the I-FGSM adversarial attack with the attack strength factor 0.1 is 0.9900 for 2C and 0.0000 for SPRITZ-1.5C, respectively. 
    
\end{itemize}

\subsection{Organization}
The study is organized as follows. In Section \ref{Background}, we provide a brief overview of the datasets and adversarial techniques employed in our study as well as some preliminary knowledge about them. 
Section \ref{Related} review relevant studies that examined the 1C property in various tools in the cybersecurity domain. In Section \ref{Methodology}, We provide the experimental setups and our proposed framework. 
Following that, we present the experimental findings in Section \ref{Results} and discuss the adversarial strategies employed against the 2C classifier (i.e., CNN). We provide a summary of our research and potential directions for further research in Section \ref{Conclusions}. 


\section{Research Background}
\label{Background}

This section outlines the datasets and adversarial techniques used in this study. First, we explain the N-BaIoT\cite{nbaiot} and RIPE-Atlas \cite{analysis} datasets. Then, we explain the adversarial attack terminology. In Table~\ref{Abbreviations}, we list all the acronyms and notations considered in our study.
\begin{table}[!h]
\centering
\caption{List of Abbreviation . \label{Abbreviations}}
\begin{tabular}{@{}|c|l|}
\hline
\textbf{Acronym} & \textbf{Description} \\ \hline

CNN & {Convolutional Neural Network} \\ \hline

RNN & {Recurrent Neural Network} \\ \hline

FL & {Federated Learning} \\ \hline

DF & {Digital Forensics} \\ \hline

2C / 1C & {2-Class Classification / 1-Class Classification} \\ \hline







PK / LK & {Perfect Knowledge / Low Knowledge} \\ \hline


CF & {Counter Forensics} \\ \hline

DL / ML & Deep Learning / Machine Learning \\ \hline

SVM & Support Vector Machine \\ \hline

GAN & Generative Adversarial Networks \\ \hline

IoT & Internet of Things \\ \hline

DDoS & Denial of Service\\ \hline

ASR & Attack Success Rate \\ \hline

Max. dist & Maximum distortion\\ \hline

PSNR & Peak Signal-to-Noise Ratio\\ \hline

$L_1$ dist & $L_1$ distance\\ \hline

FGSM & Fast-Gradient-Sign-Method\\ \hline

I-FGSM & Iterative-Fast Gradient-Sign-Method \\ \hline

JSMA  & Jacobian-based Saliency-Map-Attack\\ \hline

LBFGS  &  Limited-Memory Broyden-Fletcher-Goldfarb-Shanno\\ \hline

PGD  & Projected-Gradient-Descent\\ \hline

BIM &   Basic-Iterative-Method\\ \hline

C\&W & Carlini and Wagner \\ \hline


\end{tabular}
\end{table}

\subsection{Datasets}
\label{datasett}

We employed large real-world datasets that comprised both pristine and malicious examples to assess the effectiveness of the SPRITZ-1.5C capability in computer networks. Our study utilized the N-BaIoT and RIPE datasets to train the CNN model (2C) and auto-encoders (1Cs). This well-known dataset has been used to achieve various cybersecurity objectives lately. Employing deep auto-encoders\cite{NBaIoT_Auto}, the N-BaIoT dataset was also utilized to monitor botnet attacks in the Internet of Things (IoT) devices, IoT detection methods \cite{BaIoT_Anomaly}, and Federated Learning (FL) for identifying and mitigating IoT attacks \cite{REY2022108693}. The following paragraphs provide a detailed explanation of the dataset.\\

    \noindent \textbf{N-BaIoT dataset}: This dataset was generated using port mirroring processes to compose real and malicious traffic from nine business IoT devices. 
    The N-BaIoT dataset contains approximately seven million examples with 115 properties. As a result of the development of IoT cyber-attacks, the adversary relies on botnets to leverage such vulnerability, turning the IoT into an internet so vulnerable \cite{Turning2017, AHANGER2022108771}. 
    The N-BaIoT dataset rely on BASHLITE and Mirai botnets, which are two widely-known IoT-based botnets. These botnets perform Distributed Denial of Service (DDoS) attacks to disrupt IoT devices. 
    %
    Up to one million commercial IoT devices, principally web-connected recording devices and DVRs were impacted by BASHLITE's DDoS capabilities ~\cite{BASHLITEThreatpost}. On the other hand, the Mirai botnet has infected thousands of devices, making it among the most widely utilized IoT botnets. The Mirai botnet consists of (i)~monitoring fragile IoT devices and (ii)~DDoS attacks are used by introducing malware to vulnerable IoT devices. The two known botnets perform several attacks, including transmitting spam data, flooding UDP traffic, flooding TCP-SYN traffic, and scanning attacks. 
    \\ \\
    \noindent \textbf{RIPE-Atlas dataset}: 
    The RIPE Atlas project evaluates network packets using Internet-aware sensors \cite{tahaei2020rise}. 
    It is used to continually monitor network or client visibility from a variety of locations. Furthermore, it can perform ad-hoc connection assessments to examine the network, fix any discovered network faults, and check the DNS server availability. 
    %
    Utilizing traffic data, the RIPE dataset first attempts to determine the type of applications. Two distinct architectures are being considered to explore these datasets: Recurrent Neural Networks (RNN) and CNN. We point out that several transformation procedures were used to create the datasets \cite{analysis2}, and the flow sequence may be classified by the model for a variety of purposes.
    The model can accurately determine the majority of flow sequences. By introducing additional data into the system, the error rate is significantly reduced, even though the validity of both training and test sets exceeds 80\%.


\subsection{Adversarial Attacks}

The two types of attacks against CNN models are targeted and untargeted attacks. The untargeted attacks attempt to fool the classifier into predicting any of the inaccurate classes. In contrast, targeted attacks aim to mislead the DL model by compelling it to provide a specified target label for the adversarial example~\cite{zhang2019adversarial, aldahdooh2022adversarial}. According to the literature, the adversary can conduct attacks in three different scenarios. 
\begin{enumerate}
\item The adversary in the \textit{\textbf{white-box}} scenario has PK on the CNN (2C) by considering a specific set of adversarial examples. 
\item The adversary in the \textit{\textbf{gray-box}} scenario acquires Limited Knowledge (LK) concerning the target model. 
\item In the \textit{\textbf{black-box}} scenario, the adversary's knowledge limits the internal parameters of the two 1Cs, making this situation more realistic than the white-box scenario.
\end{enumerate}

In the following, we provide the most well-known and widely used adversarial attacks on CNN (2C) models in the DL literature, and apply them to the SPRITZ-1.5C architecture in order to evaluate its robustness. All of these strategies are applicable in both white and black-box scenarios. \\ 
%

\noindent\textit{\textbf{FGSM and I-FGSM Attacks:}} The Fast Gradient Sign Method (FGSM) was first developed as a fast sub-optimal approach in \cite{FGSM}. Given an example I, FGSM determines,
\begin{align}
\label{FGSM}
 I' =  I + \varepsilon (\max(I) - \min(I))\cdot \text{sign}(\nabla_I J_{\theta}(I,y)).
\end{align}
The attack's strength is determined by $\varepsilon$, which must be small enough to go undetectable and strong enough to induce the misclassification error. 
The FGSM method estimates the direction in which the example intensity must be modified in the first place and modifies all examples at once. It makes this by operating the gradient of the loss function. In most cases, the one-shot approach described in \eqref{FGSM} fails to create adversarial instances. By exploring the improved iterative version of the attack, a suboptimal technique may be achieved: particularly, the adversarial perturbation is generated as stated above at each iteration; hence, the revised version of the example will be as follows,
\begin{align}
\label{I-FGSM}
I_{i+1} = I_{i} + \varepsilon(\max(I_{i}) - \min(I_{i}))\cdot \text{sign}(\nabla_I J_{\theta}(I_{i},y)).
\end{align}
%

\noindent \textit{\textbf{JSMA Attack:}} This attack is optimized for the $L_0$ distance. JSMA is a greedy iterative technique that uses forward propagation to construct a saliency map at each step, indicating which samples contributed more to classifications. Large values in saliency maps imply that modifying the values in a matrix considerably reduces the chance of selecting the inaccurate class.
According to the map, the pixels are then updated one at a time by an estimated change $\theta$, which should be $\theta < 1$. The $\theta$ refers to the range of values in the scenario, and $\theta \cdot (\max(I_i) - \min(I_i))$ refers to the modification in a cell-matrix. If (such that any adversarial example can be found for a given maximum $L 0$ distortion), the operation is over when one of the following conditions is fulfilled: The attacker modifies the classifier performance, which is most likely to result in an adversarial example $I adv$ so that $l(I adv) \neq y$ and the attack alters quite so many values in a matrix. \\

\noindent\textit{\textbf{L-BFGS Attack:}} The L-BFGS method \cite{szegedy2013intriguing} is a non-linear optimization approach based on a gradient that aims to minimize perturbations to the input examples. The adversarial instances in this attack are developed by maximizing the estimated error. 
The attack is a time-consuming and computationally box-constrained optimization approach for generating adversarial examples. 
The attack optimization problem is expressed as follows for an input data $I$:
\begin{equation}
\label{Eq}
\min_{Adv_{I}} || I - Adv_{I} ||^{2}_{2}  \mbox{ subject to } F(I+r) = l.
\end{equation}
%
In the above-stated equation, $r$ is the local minimum, $l$ is the targeted label, and $F$ is the classifier. Nevertheless, the (\ref{Eq}) optimization problems are costly and problematic. The approximate solution to this problem is shown in the following using an L-BFGS box-constrained:

\begin{equation}
\label{mineqq}
\min_{Adv_{I}} t.|| I - Adv_{I} ||^{2}_{2} - J(\theta,I,Y). 
\end{equation}
This approximation's problem is to determine the minimal scalar $c$ > 0 that also fulfills the minimizer $r$ of (\ref{mineqq}). \\

\noindent \textit{\textbf{PGD Attack:}} 
This attack provides the perturbations which optimize the loss function under the specific constraints imposed by the $L^{\infty}$ distortion \cite{madry2017towards}. The example is updated to get $I_{i+1}$ at each step, after which it is projected to the examples space with a constrained $L^{\infty}$ distortion, with the maximum distortion defined to a given value $\alpha$, i.e. $I_{i+1} = \Pi_{I+\alpha}(I_{i+1})$ ($\Pi$ refer to as the projection operator). 
The standard gradient sign attack is performed with a small scale factor $\varepsilon < \alpha$; after each step, the values of each successive output are clipped to verify that the adversarial instance remains inside the $\alpha$ neighborhood of the original example. The perturbation is defined at each iteration by solving the following equation,
\begin{equation}
[I_{i+1}]_{r,c} = \text{clip}([I_{i+1}]_{r,c}, \{-\alpha,+\alpha\}).
\end{equation}
In some cases, clipping may cause this strategy to produce a suboptimal outcome. A binary search over ($\varepsilon$, $\alpha$) can be used to improve the selection of the hyperparameters. In \cite{kurakin2018adversarial}, the Basic Iterative Method refers to the PDG technique mentioned above. \\

\noindent \textit{\textbf{BIM Attack:}} The adversarial attack based on the Basic Iterative Method (BIM) is an FGSM-based improvement technique \cite{kurakin2018adversarial}. This technique iterates through the value, modifying it one step at a time and clipping the resulting value to ensure it remains within the original example's range. This method is more precise than the FGSM method.\\

\noindent \textit{\textbf{DeepFool Attack:}} The DeepFool technique intends to mislead multi-class classifiers by estimating the decision space of the classifier in order to obtain the smallest perturbations \cite{Moosavi-DezfooliDeepFool:Networks,deepfoolcite}. The minimum modification in the example required to produce an adversarial example by,
\begin{equation}   
 \delta(I,F) = \min_{r}||r||_{2} \mbox{ subject to: } F(I+r) \neq F(I). 
\end{equation}
Delta represents the consistency of F for the input X, whereas R represents the minimal perturbation. DeepFool generates examples with as small noise as possible, which should be enough to mislead the neural network architecture.\\ 

\noindent \textit{\textbf{C\&W Attack:}} 
The Carlini and Wagner (C\&W) methodology leverages high-confidence adversarial examples \cite{carlini2017towards} by providing access to the 
architecture and framework parameters. 
The C\&W attack method may be employed in three distinct distance metrics attack scenarios: $L_{2}$, $L_{0}$, and $L_{\infty}$. 
The objective of the $L_{2}$ is to specify the significance $w$ that optimizes the following expression given a benign example $I$ and a target class $t$ that differs from $I$ ($t \neq C^{*}(I) $):
\begin{equation}
 \min_{w}||\frac{1}{2}(\tanh{(w)}+1)-I||^{2}_{2} + c . f(\frac{1}{2}(\tanh{(w)}+1)).
\end{equation}
When $c$ is an acceptable constant, and $f$ is a function determined by the equation:
\begin{equation}
f(I')=max(max\{Q(I')_{i} : i \neq t\} - O(I')_{t}, -P).
\end{equation}
%
We note that $Q$ refers to the output of all layers except the softmax function and $P$ refers to the strength of the adversarial examples. The $L_2$ attack has a version defined as the $L_0$ attack.

\section{Related Works}
\label{Related}

Although the use of 1C classifiers in cybersecurity and DFs applications is not innovative, they may be found in various forensic tools. 
For example, in \cite{d2017autoencoder}, the researchers utilized 1C for video forgery identification to provide an effective system in complex environments such as social networks. The scientists employed a technique based on \textit{auto-encoders} trained on pristine data. When there is a considerable reconstruction error between both the outputs and inputs, \textit{auto-encoders} behave like 1C. 
%
One class classification (known also as 1C) is typically used for outlier detection in a variety of scenarios if a robust statistical characterization under abnormal conditions is not available. Consider the problem of recognizing acoustic diversity \cite{marchi2015novel} or predicting network intrusion \cite{xu2022improved}. Another work in \cite{perdisci2006using} provides a strategy for an adversarial intrusion detection system that integrates several 1Cs to improve the complexity of exploratory attacks. In \cite{barni2020improving}, the authors considered an ensemble model with the combination of SVM blocks to improve the security of a detector against adversarial attack. Although they achieved high security against adversarial attacks, a proposed model completely breaks against noise addition in the input.\\
\indent The \textit{Open set} problems, which have been investigated in various DF as well as security-oriented tools, are another type of 1C classification wherein LK of the domain is provided at the training phase and unidentified classes might be presented to a method during testing \cite{scheirer2012toward}, such as open set authentication of IoT \cite{huang2022novel}, incremental open set intrusion detection \cite{henrydoss2017incremental}, android malware detection \cite{sawadogo2022android}, and DDoS attack detection \cite{akgun2022new}.
%
\indent Later, 1C classifiers were integrated with Generative Adversarial Networks (GANs) to build detectors that operate on the hypothesis that there are not many malicious examples available. 
This is true for \cite{gumusbas2022ai}, which is utilized for intrusion detection systems, as well as \cite{MOTI2021102591}, which is used to identify unknown IoT infections.\\
\indent Apart from considering 1C classification in various applications for improving detection accuracy and robustness, a second approach consists of building an architecture that might resists various attacks. In \cite{9747933}, the authors demonstrated the absence of attack transferability in Computer Networks, and only a few attacks are transferable between source and target networks. As a result, they explored several deep architectures as a target network to limit attack transferability. The major disadvantage of this method is that it only responds to a few adversarial attacks. In another related work, the authors in \cite{9053318} developed a secure architecture using feature randomization to mitigate attack transferability between networks. The major disadvantage of this strategy is that it only works against a limited number of adversarial attacks, the same as previous research work. The system fails if the adversary considers an attack with a high attack parameter.
\indent In Table \ref{difference}, we provide an overview that summarizes past and present research on the various machine and deep learning applications in different applications that considers 1C classifiers. To the best of our knowledge, the security domain aspect in computer networks has not been considered in most published research in this area. We explore many application domains that used 1C classifiers. To create a model with high accuracy and security, we aim to combine the advantages of several classifiers for the first time in DL. Using prominent datasets as well as black-box scenarios, we also intend to carry out comprehensive evaluations. 
\begin{table*}[!h]
\centering
\small
\caption{Difference between our work and existing works in 1C classifier} \label{difference}
\resizebox{\textwidth}{!}{
\begin{tabular}{|l|l|l|l|l|}
\hline
\textbf{Application Domain} & \textbf{Classifier} & \textbf{Considered Datasets} & \textbf{Advantages} & \textbf{Disadvantages} \\ \hline

    Anomaly IDS \cite{perdisci2006using} & -One-Class SVM          &      -Private Dataset & -Evaded by mimicry attacks & -IDS sufficiently low                       \\ \hline
    
    Face Verification \cite{scheirer2012toward} & -One-Class SVM          &      \makecell[l]{-Caltech \\ -ImageNet} & -Improving accuracy for a face verification & -Challange in RBF kernels                         \\ \hline

    Acoustic detection \cite{marchi2015novel}& \makecell[l]{-LSTM\\-Auto-encoder}&      -PASCAL CHiME & -Denoising and Good Performance & -Dataset is not made for real events                 \\ \hline

    Video forgery detection\cite{d2017autoencoder} & -LSTM          &      -GRIP & -Considering adversarial attacks & -Fails against different manipulation     \\ \hline
    
    Intrusion detection\cite{henrydoss2017incremental} & -EVM         &      -KDDCUP’99 & -Supports incremental learning & \makecell[l]{-High soft-margin errors \\ -Not optimized EVM algorithm}                         \\ \hline
    
   \makecell[l]{Multimedia Security  \cite{9053318}} & \makecell[l]{-CNN\\-SVM}          &      -RAISE & -Improve a security & -Works against few attacks                         \\ \hline
    
    \makecell[l]{Multimedia Forensics  \cite{barni2020improving}} & -SVM         &      -RAISE & -Improve a security & -Sensitive to noise                          \\ \hline
    
    Internet of Things \cite{MOTI2021102591}  & \makecell[l]{-CNN, GAN\\-LSTM }        &     \makecell[l]{-Detect unseen \\ IoT malware} & \makecell[l]{-Higher detection rate \\ -Can be used in settings with \\insufficient or imbalanced data}  & -Failed to detect adversarial examples                        \\ \hline
   
   Network Intrusion Detection  \cite{xu2022improved} & -Bi-GAN          &      \makecell[l]{-NSL-KDD \\ -CIC-DDoS2019} & -Less training overhead &  -High false positive rate                          \\ \hline
   
   Open set \cite{huang2022novel} & \makecell[l]{-CNN\\-OpenMAX}          &      -ADS–B & -Applicable on real world problems & -Low robustness of the features                          \\ \hline
   
   Android malware detection \cite{sawadogo2022android} & -Different classifiers  &      -Private Dataset & -Increase high detection rate & -Sensitive to adversarial attacks                         \\ \hline
   
   Intrusion detection \cite{akgun2022new}  & \makecell[l]{-DNN\\-CNN\\-LSTM}  &      -CIC-DDoS2019 & \makecell[l]{-Good inference time \\ -Smaller number of trainable parameters}  & -Sensitive to adversarial attacks                          \\ \hline
   
   Computer Networks \cite{9747933}  & -CNN  &      \makecell[l]{-N-BaIoT \\ -DGA \\ -RIPE Atlas} & -Considering adversarial attacks  & -Responds to a few adversarial attacks                         \\ \hline
   
   This study  & \makecell[l]{-CNN\\-Auto-encoder}  &      \makecell[l]{-N-BaIoT\\ -RIPE } & -\makecell[l]{-Improve a Computer Networks security\\ -Robust against noise addition }  & \makecell[l]{-Need to investigate against causative \\ attacks (future research work)}   \\ \hline
   
\end{tabular}
}
\end{table*}

\section{Evaluation Methodology}
\label{Methodology}

In this section, we specifies the detection task. Then, we discuss the SPRITZ-1.5C framework, including adversarial techniques on the 2C and 1C 
as well as the SPRITZ-1.5C classifier's training process. 
In this study, we considered two class classifications for pristine and malicious examples: hypothesis H0 refers to the case of pristine examples provided and without any additional processing. In contrast, hypothesis H1 refers to the case of modified or altered examples. In this case, in a real-world scenario, the attacker is constantly eager to apply adversarial attacks to H1 to prevent a correct detection. In addition, we suppose that the attacker's purpose is to evade detection of the manipulation, i.e., to cause a missed detection error. In CF, the most standard attack is an integrity violation attack~\cite{dasgupta2022machine}. With $P_{MD}$ and $P_{FA}$, respectively, we can represent the probabilities of a missed detection error, which is the possibility that a malicious example will be misidentified for a clean example, and a false alarm probability, which is the possibility that a clean example will be wrongly identified as a tampered example. 

\subsection{Architecture of the SPRITZ-1.5C classifier}
\begin{figure}[!h]
\begin{center}
\includegraphics[width=0.48\textwidth]{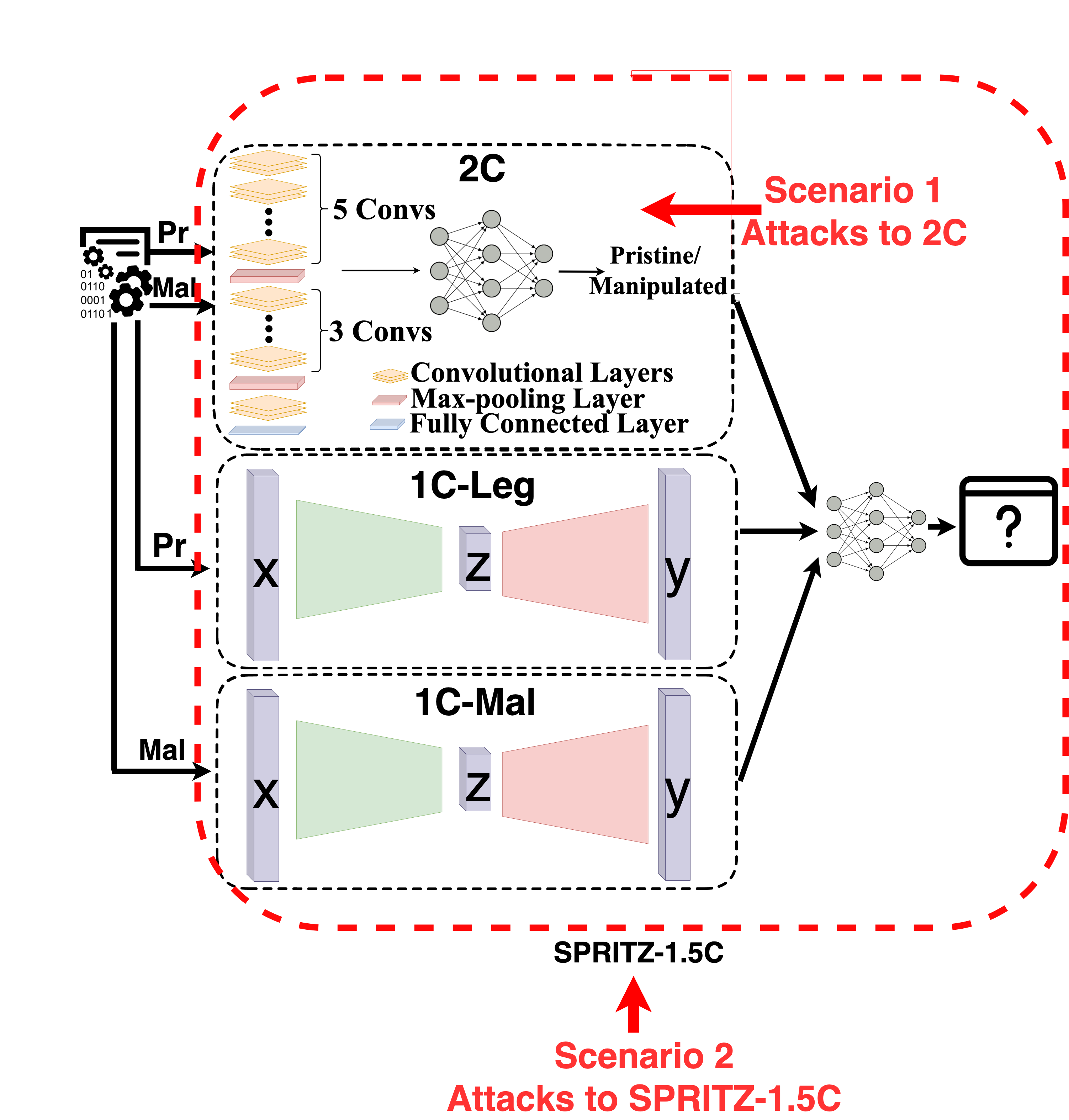}
\caption{The framework of the SPRITZ-1.5C classifier employed in this study is as follows. The input data contains both pristine and malicious examples. In this diagram, we considered CNN architecture as a 2C, taking into account auto-encoder "1C-Leg" trained with pristine examples, auto-encoder "1C-Mal" trained with malicious examples, and final dense classifier, also known as a classifier combination. As a consequence, the outputs of these three classifiers are employed to train a dense classifier.}
\label{Ensemble}
\end{center}
\end{figure}

The SPRITZ-1.5C architecture used in this study by considering the ensemble method is shown in Figure \ref{Ensemble}. 
%
Three classifiers were trained in parallel using dataset examples: a 2C (here, CNN) trained with pristine and malicious instances from both classes from the N-BaIoT and RIPE-Atlas datasets independently, and two 1Cs (here, two auto-encoders), one trained with pristine and the other with malicious examples. The results of these classifiers are subsequently processed by a final dense classifier, which makes a final decision. \\
As stated in the introduction, in the absence of attacks, 2C classification techniques can achieve high accuracy; nevertheless, they do not generalize properly to examples that were improperly represented during the training process, enabling the adversary to carry out his attack by exploiting the unexplored areas of the features space. 
%
This scenario is illustrated in Figure \ref{figRegion:a}, where the adversary takes full advantage of the availability of empty spaces to induce a missed detection error, exposing the 2C vulnerable to attacks. However, 1C are inherently more robust to adversarial attacks since they provide a closed region that only includes examples through a specific class, commonly called the H0 class. Figure. \ref{figRegion:b} illustrates this impact, showing that a larger distortion is required to move an example from the H1 (malicious examples) region to the H0 (pristine examples) region.
%
As a consequence, the acceptance region in Figure ~\ref{figRegion:c} (referred to as SPRITZ-1.5) has comparable robustness and performance against adversarial attacks as those obtained by the 1C (see Figure ~\ref{figRegion:b}) and 2C classifier (see Figure~\ref{figRegion:c}). 
To clarify more regarding Figure \ref{Ensemble}, we present two possible scenarios:
\begin{itemize}
    \item \textbf{Scenario 1:} The attacker exploits several adversarial attacks against the 2C classifier. Afterward, we analyze the performance of all classifiers using adversarial examples generated by the 2C classifier.
    \item \textbf{Scenario 2:} In this scenario, the attacker performs a variety of adversarial attacks against the entire secure SPRITZ-1.5C architecture.
\end{itemize}

\subsection{Network Architecture}
\begin{figure}[!h]
\begin{center}
\includegraphics[width=0.47\textwidth]{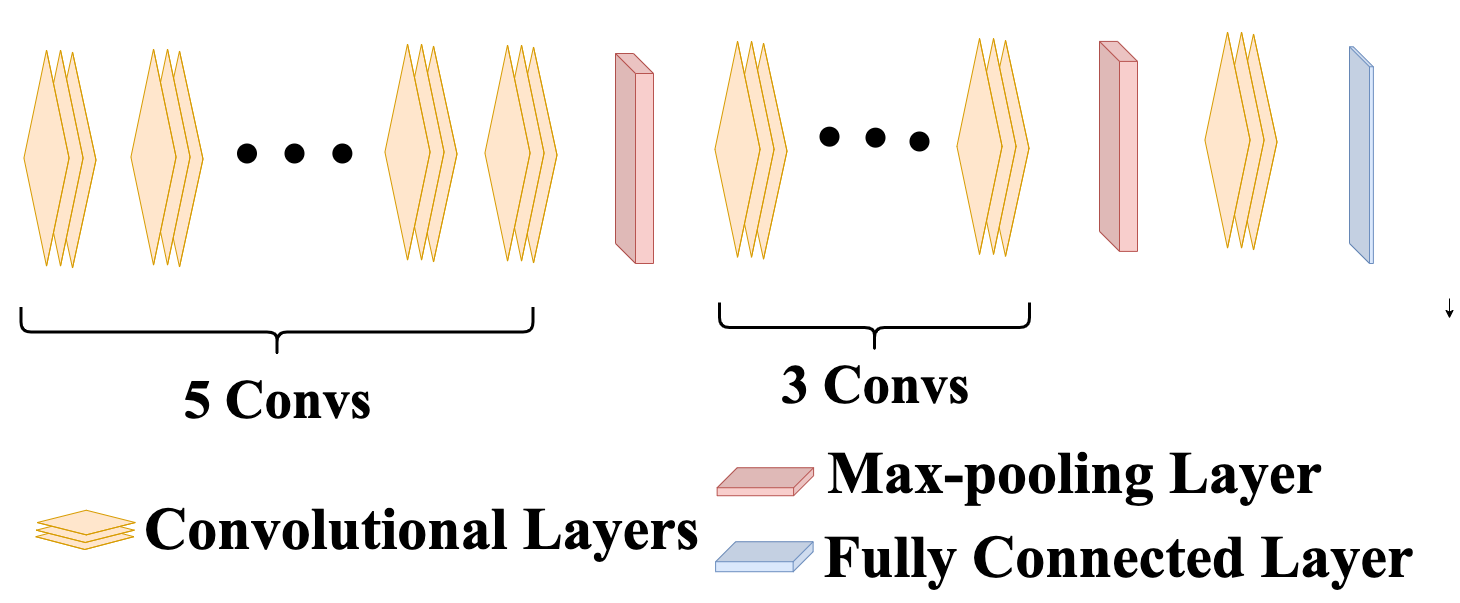}
\caption{The proposed 2C (CNN) architecture's pipeline}
\label{2CNet}
\end{center}
\end{figure}

We adopted the network configuration provided by \cite{8451698} for the 2C classifier. 
This network has nine convolutional layers followed by max-pooling layers, and a dense layer followed by a sigmoid layer. The 2C architecture's process is depicted in Figure \ref{2CNet}. For all convolutions, we employ a kernel size of $3\times3$ and a stride of 1, and for max-pooling with a stride 2 with a kernel size of $2\times2$. Only one dense layer is employed, which reduces the number of parameters.
The network has several convolutions to transmit every neuron before the first max-pooling layer effectively. Adjusting the stride to one, the best spatial information will remain. We employ this network configuration because it offers a high level of accuracy over training. 
%

\begin{figure}[!h]
\begin{center}
\includegraphics[width=0.45\textwidth]{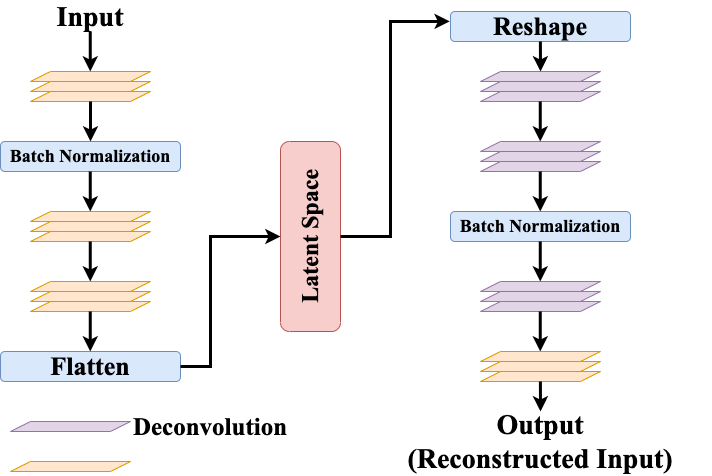}
\caption{The proposed 1C (auto-encoders) architecture's pipeline.}
\label{autoNet}
\end{center}
\end{figure}

For 1C-Leg and 1C-Mal auto-encoders, we employed a simple CNN-based auto-encoder. We used convolutional layers, batch normalization layers, and a fully-connected layer for the encoding part. For decoding, we adopted a fully-connected layer, de-convolutional~(transposed convolutions) layers, a batch normalization layer, and a convolutional layer. In addition, we have a latent space between the encoder and the decoder, which has just one fully-connected layer. The latent space is the output of the encoder and is used as input for the decoder. In the encoder, the network has some convolutions to capture the most important features of the input samples. For decreasing the size of feature maps, the strides of all convolutions were set to two. To build a compressed representation of input samples, auto-encoders can be utilized. For this reason, we exploit this network architecture because it allowed us to have extremely good similarity for the original input and network output, i.e., reconstructed input using latent space features. The 1C model is illustrated in Figure \ref{autoNet}. We set $3\times3$ kernel sizes and strides of two for all convolutions and deconvolutions. For the latent space, we considered a dense layer with a size of 512 that may help us to increase the effectiveness of auto-encoders in the SPRITZ-1.5C classifier.\\
Three dense layers are considered to perform the classification task for the dense network. The first dense layer is created by concatenating the extracted flatten layer of the 2C classifier with the size of 1728 to latent spaces, i.e., the output of the encoder part of two 1C classifiers that size of each one is 512. The concatenation task will generate a dense layer with a size of 2752 neurons. As we have two classes (pristine and malicious), layers of size 128 and 2 are considered for the second and third dense layers, respectively, which are suitable for our classification task.\\

\subsection{Experimental Setup}

We employed 29000 samples from the N-BaIoT dataset for training, 10000 for verification, and 10000 for testing, per class to train a 2C classifier. We employed the same dataset split strategy for training 1C-Leg, where only pristine examples were considered, while only malicious examples were considered for 1C-Mal. The dense classifier (also known as a cmb-classifier) is fed by the concatenation of a flattening layer from a 2C and two latent layers from 1C-Leg and 1C-Mal. In this regard, the 2C flatten layer size is 1728, the latent layer size from the 1C-Leg is 512, and the 1C-Mal is also 512. As a consequence, the input size fed to the cmb-classifier is 2752. Concerning the RIPE dataset, we considered 30000 examples for training a 2C classifier, 10000 for validation, and 10000 for testing, per class. We utilized the same dataset split strategy for training 1C-Leg that we employed for N-BaIoT, where only pristine examples were considered, whereas only malicious examples were considered for 1C-Mal. 
Furthermore, the cmb-classifier size is 2752, the same as the N-BaIoT cmb-classifier.

\begin{figure}[!h]
\begin{center}
\includegraphics[width=0.40\textwidth]{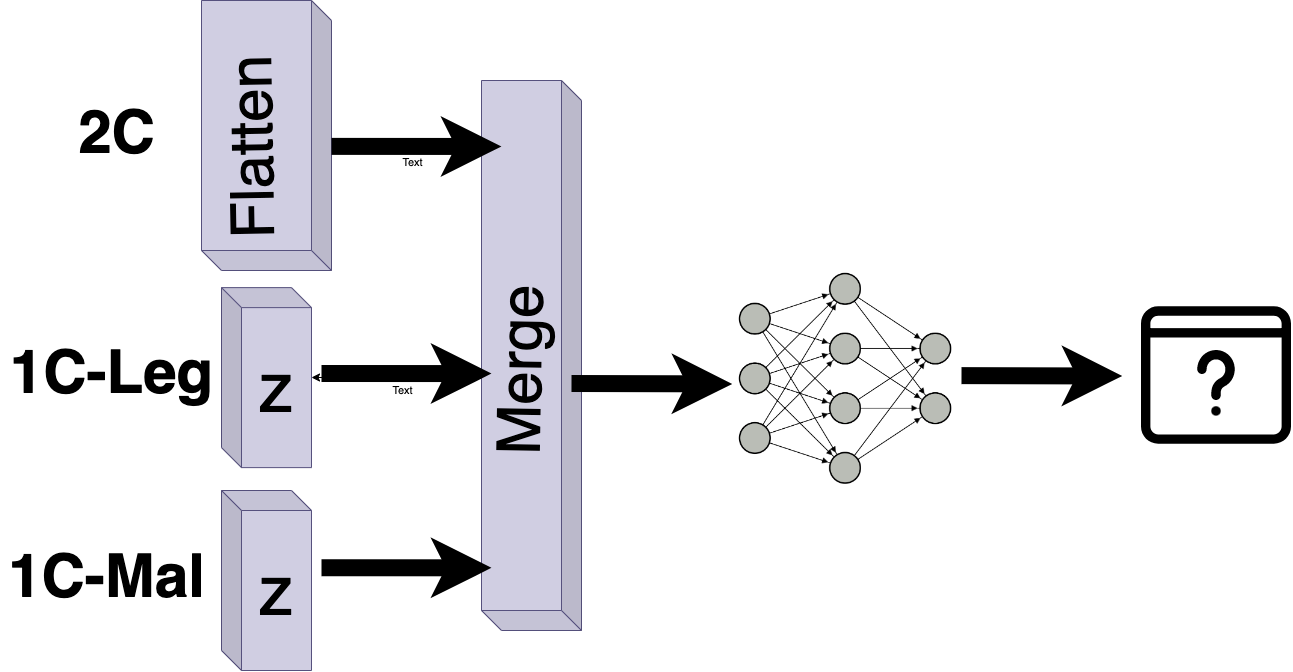}
\caption{The cmb-classifier training pipeline.}
\label{DensePipe}
\end{center}
\end{figure}

%
To clarify the training of the cmb-classifier, we utilized the features from the flattening layer (2C) and the features from the latent space of auto-encoders (1Cs). Figure \ref{DensePipe} exhibits the training pipeline considered for the cmb-classifier.
%
All features are then merged and structured into a dense network. In all cases, the input size is always $64\times64$. We believe that this set of features is sufficient to generalize a CNN, auto-encoders, and a dense network. 
We employed TensorFlow and Python to create our classifiers, using the Keras API. 
In our experiments, we employed the Intel CoreTM i7 processor 10750H, the GeForce NVIDIA 2060 RTXTM with GDDR6 6GB GPU, and DDR4 32GB, and Ubuntu operating system version 20.04 were used in our tests. 
We provided the Python code for the simulation on Github \cite{SPRITZ15C}. For SPRITZ-1.5C, 500 training cycles have been carried out across all classifiers. We used the Adam solution with a momentum of 0.99 and a learning rate ($lr$= $10^{-4}$). For training and validation, the batch size for the N-BaIoT and RIPE datasets is 16.
\subsection{Security assessment}
The SPRITZ-1.5C classifier's security is evaluated by examining the study's validity under adversarial attacks. 
Compared to those achieved by 2C under the same attacks, these outcomes are close to each other. To prove that the SPRITZ-1.5C model provides better security than the 2C classifier, this study aims to demonstrate that system compromise causes more distortion in the scenarios that are being attacked. 
We especially took into account the threat model below: \\
\begin{itemize}
    \item The \textbf{attacker's objective} is to alter a malicious example so that the feature representation is transferred into the pristine area, leading to a missed detection error. 
    \item By using terminology from \cite{biggio2013evasion}, we highlight a PK where the adversary has full knowledge of the model.
    This refers to taking into account the \textbf{attacker's knowledge} scenario. 
    \item In terms of \textbf{attacker capabilities}, we concentrate on exploratory attacks \cite{hernandez2022adversarial}, that is, attacks conducted during the testing phase. This category comprises the majority of the CF approaches proposed in the literature.
\end{itemize}

\section{Experimental Results and Discussion}
\label{Results}
We present experimental results in 2C and SPRITZ-1.5C separately in this section. First, we present the accuracy rate of all classifiers in the absence of attacks. Then, we conduct several adversarial attacks against 2C and evaluate the performance of each classifier. Finally, we employed the same adversarial attacks on SPRITZ-1.5C in a black-box setting to understand its robustness fully. All experiments were carried out using two datasets, N-BaIoT and RIPE. 

\subsection{Experimental Results}
Table \ref{Test_Acc_Classifiers} provides the test accuracy values of the 2C, 1C pristine, 1C malicious, and the combination dense classifier (acronym to cmb-classifier). We see that the performance of the cmb-classifier increase slightly compared to 2C.
\begin{table}[!h]
\scriptsize
\centering
\caption{Test accuracy values of all the classifiers. The performance of the SPRITZ-1.5C systems are those reported for cmb-classifier. \label{Test_Acc_Classifiers}}
\begin{tabular}{|c|c|c|c|c|}
\hline
\textbf{Dataset} & \textbf{2C} & \textbf{1C-Leg} & \textbf{1C-Mal} & \textbf{cmb-classifier} \\ \hline
\rowcolor{grey}
N-BaIoT & 99.98\% & 98.86\% & 98.80\% & 100\% \TBstrut\\ \hline
RIPE & 99.76\% & 98.32\% & 98.51\% & 100\% \TBstrut\\ \hline
\end{tabular}
\end{table}
The average ASR on 2C based on CNN networks and dense networks utilized in SPRITZ-1.5C are represented in the results when we applied different adversarial attacks. In addition, we took into account the average PSNR on 500 samples from the test folder, the average L1 distortion ($L_1$ dist), and the average maximum absolute distortion (Max. dist) as other metrics. In addition, we provide running attack timings in seconds while performing adversarial assaults on CNN and dense networks.  \\
%
\indent For the I-FGSM attacks, we considered the strength attack factors $\varepsilon$ to 0.1, 0.01, and 0.001, and we employed the same attack parameters for the FGSM. We fixed the number of steps $S$ to 10. The strength factor $\theta$ for JSMA adversarial attack is set at 0.1 and 0.01. In the case of other attacks such as DeepFool, LBFGS, BIM, and PGD, we considered the default attack variable, which we believe is sufficient to fool a 2C network. Finally, we looked into C\&W adversarial attacks with various confidence factors such as 0, 50, and 100. \\

\subsubsection{\textbf{Performance under attacks}} 
In this part, we evaluate the effectiveness of the 2C and SPRITZ-1.5C in the face of attacks. We started by deploying all attacks on 2C and then evaluated the SPRITZ-1.5C to see how it performed. Subsequently, to evaluate the security level of SPRITZ-1.5C, we conducted adversarial attacks directly on it. \\

\textbf{Attack against 2C (Scenario 1):} For detecting task N-BaIoT and RIPE pristine examples from malicious examples, we first conduct the different adversarial attacks against 2C to demonstrate that this classifier is intrinsically sensitive to all adversarial attacks. 
As predicted, the attack was consistently successful in producing an incorrect classification, and after running, all malicious examples are classified as pristine. Table \ref{2C_Attack}, and \ref{2C_Attack_RIPE} reported when the eight different adversarial attacks were employed against the 2C classifier and when the models were trained using the N-BaIoT and RIPE-Atlas datasets.
The findings in this table relate to the average ASR  on 2C, average PSNR on 500 examples, average $L 1$ distortion, average maximum absolute distortion (Max. dist), and average running attacks time in seconds, as was stated earlier. The adversarial attack was effective if the average ASR threshold was higher than 50\%. \\
\begin{table}[!h]
\scriptsize
\centering
\caption{EXPERIMENTAL RESULTS FOR ADVERSARY ATTACKS ON A 2C (Trained with N-BaIoT Dataset). The time of running attacks on the models is reported in seconds in this table. \label{2C_Attack} }
\begin{tabular}{|c|c|c|c|c|c|}
\hline

\textbf{Attack Type} & \textbf{PSNR} & \textbf{$L_1$ dist} & \textbf{Max. dist} & \textbf{ASR} & \textbf{Exe. Time} \\ \hline
\rowcolor{grey}
I-FGSM, $\varepsilon$= 0.1 & 17.617 & 29.691 & 40.733 & 0.9900 & 315.518 \TBstrut\\ \hline
I-FGSM, $\varepsilon$= 0.01 & 17.996 & 28.557 & 39.101 & 0.9500 & 3044.2015 \TBstrut\\ \hline
\rowcolor{grey}
I-FGSM, $\varepsilon$= 0.001 & 17.997 & 28.410 & 38.927 & 0.9463 & 29805.6208 \TBstrut\\ \hline
FGSM, $\varepsilon$= 0.1 & 17.277 & 30.856 & 40.712 & 0.9800 & 29.6365 \TBstrut\\ \hline
\rowcolor{grey}
FGSM, $\varepsilon$= 0.01 & 17.414 & 30.396 & 40.098 & 0.9756 & 46.8270 \TBstrut\\ \hline
FGSM, $\varepsilon$= 0.001 & 17.546 & 29.931 & 39.470 & 0.9500 & 152.3631 \TBstrut\\ \hline
\rowcolor{grey}
JSMA, $\theta$= 0.1 & 17.432 & 7.025 & 178.500 & 0.9900 & 3513.4855 \TBstrut\\ \hline
JSMA, $\theta$= 0.01 & \textbf{Fails} & \textbf{Fails} & \textbf{Fails} & \textbf{Fails} & \textbf{Fails} \TBstrut\\ \hline
\rowcolor{grey}
DeepFool, Default & 23.489 & 7.162 & 175.386 & 1.0000 & 221.2673 \TBstrut\\ \hline
LBFGS, Default & 24.145 & 7.143 & 132.766 & 0.9700 & 2825.4893 \TBstrut\\ \hline
\rowcolor{grey}
BIM, Default & 17.905 & 28.823 & 38.980 & 0.9800 & 559.6410 \TBstrut\\ \hline
PGD, Default & 17.915 & 28.876 & 38.747 & 0.9800 & 2279.5689 \TBstrut\\ \hline
\rowcolor{grey}
C\&W, $c$ = 0 & 24.130 & 6.981 & 131.851 & 1.0000 & 9012.3253 \TBstrut\\ \hline
C\&W, $c$ = 50 & 23.60 & 7.099 & 127.714 & 1.0000 & 7968.1794 \TBstrut\\ \hline
\rowcolor{grey}
C\&W, $c$ = 100 & 23.50 & 7.100 & 127.8402 & 1.0000 & 7464.7506 \TBstrut\\ \hline
\end{tabular}
\end{table}
\begin{table}[!h]
\scriptsize
\centering
\caption{EXPERIMENTAL RESULTS FOR ADVERSARY ATTACKS ON A 2C (Trained with RIPE Dataset). The time of running attacks on the models is reported in seconds in this table. \label{2C_Attack_RIPE}}
\begin{tabular}{|c|c|c|c|c|c|}
\hline
\textbf{Attack Type} & \textbf{PSNR} & \textbf{$L_1$ dist} & \textbf{Max. dist} & \textbf{ASR} & \textbf{Exe. Time} \\ \hline
\rowcolor{grey}
I-FGSM, $\varepsilon$= 0.1 & 16.415 & 27.541 & 40.883 & 0.9900 & 423.618 \TBstrut\\ \hline
I-FGSM, $\varepsilon$= 0.01 & 16.763 & 26.998 & 40.765 & 0.9863 & 5440.1006 \TBstrut\\ \hline
\rowcolor{grey}
I-FGSM, $\varepsilon$= 0.001 & 16.761 & 26.986 & 39.601 & 0.9556 & 39805.1309 \TBstrut\\ \hline
FGSM, $\varepsilon$= 0.1 & 16.107 & 28.733 & 38.510 & 0.9900 & 35.5140 \TBstrut\\ \hline
\rowcolor{grey}
FGSM, $\varepsilon$= 0.01 & 16.212 & 28.190 & 38.190 & 0.9856 & 56.9280 \TBstrut\\ \hline
FGSM, $\varepsilon$= 0.001 & 16.440 & 27.721 & 36.255 & 0.9800 & 250.1616 \TBstrut\\ \hline
\rowcolor{grey}
JSMA, $\theta$= 0.1 & 16.322 & 7.125 & 187.652 & 0.9832 & 4510.3905 \TBstrut\\ \hline
JSMA, $\theta$= 0.01 & \textbf{Fails} & \textbf{Fails} & \textbf{Fails} & \textbf{Fails} & \textbf{Fails} \TBstrut\\ \hline
\rowcolor{grey}
DeepFool, Default & 39.515 & 8.252 & 180.165 & 1.0000 & 301.1695 \TBstrut\\ \hline
LBFGS, Default & 29.450 & 8.120 & 143.855 & 0.9800 & 39215.4990 \TBstrut\\ \hline
\rowcolor{grey}
BIM, Default & 20.905 & 29.763 & 40.129 & 0.9900 & 439.5009 \TBstrut\\ \hline
PGD, Default & 20.816 & 30.521 & 40.556 & 0.9700 & 3289.4699 \TBstrut\\ \hline
\rowcolor{grey}
C\&W, $c$ = 0 & 22.140 & 5.871 & 132.751 & 1.0000 & 8212.5360 \TBstrut\\ \hline
C\&W, $c$ = 50 & 20.41 & 6.098 & 136.604 & 1.0000 & 7968.2414 \TBstrut\\ \hline
\rowcolor{grey}
C\&W, $c$ = 100 & 20.52 & 6.510 & 136.6512 & 1.0000 & 7464.7706 \TBstrut\\ \hline
\end{tabular}
\end{table}
%
%
The I-FGSM attack has a predefined number of steps $S$ for attacks, which is 10. where SSS, which refers to search step size and is related to the optimum strength of attacks; 
therefore, the optimal strength is considered in the [$\epsilon_{s}$: 0.1, 0.01, and 0.001] range. A powerful attack is often considered when setting a higher $\epsilon_{s}$, but we need to apply high distortion in the input example; therefore, we will have high PSNR. In our studies, we examined $\epsilon_{s}$ = 0.1, 0.01, and 0.001, which mislead a 2C architecture also that average PSNR remains around 17db, implying that attacks must apply higher distortion. This attack scenario also happens for the FGSM; even though there are no steps in this attack, it can mislead a network with a close average PSNR, similar to I-FGSM. The BIM technique is an enhancement of the FGSM approach. The approach is to iteratively adjust the value one step at a time and prune the resulting value to verify that it is within the prescribed range of the original sample. Regarding BIM, we investigated using a default parameter since the default parameter is sufficient to deceive the 2C classifier. PGD adversarial attack, similar to FGSM, is concerned with determining the perturbation that maximizes the loss function under given $l$ distortion constraints. In terms of PGD, we considered the default parameter, which we believe is sufficient to mislead a 2C architecture.

The JSMA's parameter T is set at 7. The method's number of iterations is set to 2000 by default. The relative alternation per pixel is fixed to 0.01 and 0.1. We did not examine parameters lower than 0.01 since JSMA already fails to deceive a 2C architecture with a parameter of 0.01.
We utilized the default attack setting for L-BFGS, which attempts to estimate the optimal solutions of the optimization method that the attack should handle to determine the lowest perturbation in an adversarial example that causes the error in predictions (typical gradient-descent approach).

In DeepFool, adversarial attacks initialize the sample confined by a classifier's decision boundaries. The area of the decision boundaries determines the sample's class designation. A small vector is conducted with a sample approximated by the polyhedron's boundary in each cycle. 
The perturbations are then carried out to an example in each iteration to determine the overall perturbation based on initial decision classifier limitations. 
Regarding C\&W adversarial attack, C\&W claims that a series of attacks evaluate norm-restricted additive perturbations and completely destroy defensive distillation. It is further shown that when the perturbation is produced using an exposed white-box model, their attack effectively deceives a network that has been defensively distilled under black-box settings. The results in Table \ref{2C_Attack} and Table \ref{2C_Attack_RIPE} show that both attacks can mislead a 2C architecture. In reality, we investigated both attacks in this study since recent studies show that both are among the most powerful attacks in DL. \\

\textbf{Performance of 2C attacks against SPRITZ-1.5C and Auto-encoders:} In the previous section, we conducted various attacks on the 2C classifier and concluded that most attacks fully fooled a detector. After performing attacks to 2C with 500 samples, we kept attack samples separately to test SPRITZ-1.5C since we wanted to know if SPRITZ-1.5C could determine whether these samples were attack samples. The results of 2C attack samples against the SPRITZ-1.5C are shown in Table \ref{Test_Nets} for the N-BaIoT dataset and Table \ref{Test_Nets_RIPE} for a RIPE dataset. We observe that the attacked samples we achieved against 2C are ineffective since most of the SPRITZ-1.5C test accuracy is close to 0\%. \\
\indent Additionally, the architecture of SPRITZ-1.5C includes two auto-encoders, one of which is trained with pristine samples while the other is trained with malicious samples (see Figure \ref{Ensemble}). For ease of use, we refer to auto-encoders trained on pristine samples as 1C-Leg and those trained on malicious samples as 1C-Mal. Given that we obtained the adversarial samples from the 2C architecture, we tested 1C-Leg and 1C-Mal against these attacks. 
Attacks over 1C-Leg and 1C-Mal are ineffective, demonstrating that using a constrained acceptance area makes the 1C more resistant to adversarial attacks (see Figure \ref{figRegion:b}). As a result, the attack is inefficient in causing an inaccurate classification for the SPRITZ-1.5C (see Figure \ref{figRegion:c}) but efficient in causing a missed classification error in 2C (see Figure \ref{figRegion:a}).
%
\begin{table}[!h]
\scriptsize
\centering
\caption{The percentage of attacked cases that were misclassified. The attack is directed employed at 2C (N-BaIoT Dataset). \label{Test_Nets}}
\begin{tabular}{|c|c|c|c|}
\hline
\textbf{Attack Type} & \textbf{1C-Leg} & \textbf{1C-Mal} & \textbf{SPRITZ-1.5C}  \TBstrut\\ \hline
\rowcolor{grey}
I-FGSM, $\varepsilon$ = 0.1 & 0.0020 & 0.0100 & 0.0000  \TBstrut\\ \hline
I-FGSM, $\varepsilon$ = 0.01 & 0.0001 & 0.0020 & 0.0000  \TBstrut\\ \hline
\rowcolor{grey}
I-FGSM, $\varepsilon$ = 0.001 & 0.0000 & 0.0000 & 0.0000  \TBstrut\\ \hline 
FGSM, $\varepsilon$ = 0.1 & 0.0050 & 0.0010 & 0.0000  \TBstrut\\ \hline
\rowcolor{grey}
FGSM, $\varepsilon$ = 0.01 & 0.0001 & 0.0030 & 0.0000  \TBstrut\\ \hline
FGSM, $\varepsilon$ = 0.001 & 0.0000 & 0.0000 & 0.0000  \TBstrut\\ \hline
\rowcolor{grey}
JSMA, $\theta$ = 0.1 & 0.0011 & 0.0150 & 0.0000  \TBstrut\\ \hline
JSMA, $\theta$ = 0.01 & \textbf{Fails} & \textbf{Fails} & \textbf{Fails}  \TBstrut\\ \hline
\rowcolor{grey}
DeepFool, Default & 0.1812 & 0.1200 & 0.0000  \TBstrut\\ \hline
LBFGS, Default & 0.0010 & 0.0002 & 0.0000  \TBstrut\\ \hline
\rowcolor{grey}
BIM, Default & 0.0050 & 0.0001 & 0.0000  \TBstrut\\ \hline
PGD, Default & 0.0005 & 0.0000 & 0.0000  \TBstrut\\ \hline
\rowcolor{grey}
C\&W, $c$ = 0 & 0.0004 & 0.0000 & 0.0001  \TBstrut\\ \hline
C\&W, $c$ = 50 & 0.0018 & 0.0021 & 0.0061  \TBstrut\\ \hline
\rowcolor{grey}
C\&W, $c$ = 100 & 0.1012 & 0.1001 & 0.0070  \TBstrut\\ \hline
\end{tabular}
\end{table}
%
%
\begin{table}[!h]
\scriptsize
\centering
\caption{The percentage of attacked cases that were misclassified. The attack is directed employed at 2C (RIPE Dataset). \label{Test_Nets_RIPE}}
\begin{tabular}{|c|c|c|c|}
\hline
\textbf{Attack Type} & \textbf{1C-Leg} & \textbf{1C-Mal} & \textbf{SPRITZ-1.5C}  \TBstrut\\ \hline
\rowcolor{grey}
I-FGSM, $\varepsilon$ = 0.1 & 0.0040 & 0.0031 & 0.0000  \TBstrut\\ \hline
I-FGSM, $\varepsilon$ = 0.01 & 0.0006 & 0.0010 & 0.0000  \TBstrut\\ \hline
\rowcolor{grey}
I-FGSM, $\varepsilon$ = 0.001 & 0.0000 & 0.0000 & 0.0000  \TBstrut\\ \hline 
FGSM, $\varepsilon$ = 0.1 & 0.0300 & 0.0020 & 0.0000  \TBstrut\\ \hline
\rowcolor{grey}
FGSM, $\varepsilon$ = 0.01 & 0.0020 & 0.0010 & 0.0000  \TBstrut\\ \hline
FGSM, $\varepsilon$ = 0.001 & 0.0000 & 0.0000 & 0.0000  \TBstrut\\ \hline
\rowcolor{grey}
JSMA, $\theta$ = 0.1 & 0.0800 & 0.0030 & 0.0000  \TBstrut\\ \hline
JSMA, $\theta$ = 0.01 & \textbf{Fails} & \textbf{Fails} & \textbf{Fails}  \TBstrut\\ \hline
\rowcolor{grey}
DeepFool, Default & 0.0240 & 0.0120 & 0.0030  \TBstrut\\ \hline
LBFGS, Default & 0.0300 & 0.0010 & 0.0000  \TBstrut\\ \hline
\rowcolor{grey}
BIM, Default & 0.0050 & 0.0001 & 0.0000  \TBstrut\\ \hline
PGD, Default & 0.0009 & 0.0010 & 0.0000  \TBstrut\\ \hline
\rowcolor{grey}
C\&W, $c$ = 0 & 0.0081 & 0.0030 & 0.0010  \TBstrut\\ \hline
C\&W, $c$ = 50 & 0.0041 & 0.0033 & 0.0051  \TBstrut\\ \hline
\rowcolor{grey}
C\&W, $c$ = 100 & 0.2021 & 0.2201 & 0.0084  \TBstrut\\ \hline
\end{tabular}
\end{table}
%

According to Table \ref{Test_Nets} and Table \ref{Test_Nets_RIPE}, attacking 2C is insufficient to deceive the SPRITZ-1.5C classifier as well as both 1C classifiers. According to the values in these tables, the misclassification rate in SPRITZ-1.5C is 0\% in most adversarial attacks, and only in C\&W, the misclassification rates are 0.0001, 0.0061, and 0070. To better understand, for the I-FGSM adversarial attack in a Table \ref{Test_Nets} with a parameter of 0.1, we evaluate all classifiers using attack examples given by 2C; thus, testing 1C-Leg, the misclassification rate is 0.0020, 1C-Mal is 0.0100, and SPRITZ-1.5C is 0.0000. Furthermore, we have seen similar behavior when SPRITZ-1.5C architecture is trained on a RIPE dataset. As a result, as shown in Table \ref{Test_Nets_RIPE}, we assess all classifiers using attack examples provided by 2C; so, evaluating 1C-Leg, the misclassification rate is 0.0040, 1C-Mal is 0.0031, and SPRITZ-1.5C is 0.0000.
Thereby, we can state that 1C-Leg and 1C-Mal enhance SPRITZ-1.5C in improving security. \\

\textbf{Attack against the SPRITZ-1.5C (Scenario 2):} We have launched adversarial attacks against 2C and evaluated the misclassification rate of the 1C-Leg, 1C-Mal, and SPRITZ-1.5C. We have discovered that attacks in 2C are insufficient to mislead the entire architecture. 
In this scenario, we will analyze the security of the SPRITZ-1.5C model to determine if the attacker can deceive this model or become fail.
\begin{table}[!h]
\scriptsize
\centering
\caption{EXPERIMENTAL RESULTS FOR ADVERSARY ATTACKS ON A SPRITZ-1.5C (Trained with N-BaIoT Dataset). The time of running attacks on the models is reported in seconds in this table.  \label{SPRITZ_Attack}}
\begin{tabular}{|c|c|c|c|c|c|}
\hline
\textbf{Attack Type} & \textbf{PSNR} & \textbf{$L_1$ dist} & \textbf{Max. dist} & \textbf{ASR} & \textbf{Exe. Time} \\ \hline
\rowcolor{grey}
I-FGSM, $\varepsilon$ = 0.1 & \textbf{Fails} & \textbf{Fails} & \textbf{Fails} & \textbf{Fails} & 406.415 \TBstrut\\ \hline
I-FGSM, $\varepsilon$ = 0.01 & \textbf{Fails} & \textbf{Fails} & \textbf{Fails} & \textbf{Fails} & 4144.1005 \TBstrut\\ \hline
\rowcolor{grey}
I-FGSM, $\varepsilon$ = 0.001 & \textbf{Fails} & \textbf{Fails} & \textbf{Fails} & \textbf{Fails} & 37605.4521 \TBstrut\\ \hline
FGSM, $\varepsilon$ = 0.1 & \textbf{Fails} & \textbf{Fails} & \textbf{Fails} & \textbf{Fails} & 49.4734 \TBstrut\\ \hline
\rowcolor{grey}
FGSM, $\varepsilon$ = 0.01 & \textbf{Fails} & \textbf{Fails} & \textbf{Fails} & \textbf{Fails} & 66.8612 \TBstrut\\ \hline
FGSM, $\varepsilon$ = 0.001 & \textbf{Fails} & \textbf{Fails} & \textbf{Fails} & \textbf{Fails} & 451.2393 \TBstrut\\ \hline
\rowcolor{grey}
JSMA, $\theta$ = 0.1 & \textbf{Fails} & \textbf{Fails} & \textbf{Fails} & \textbf{Fails} & 4313.7812 \TBstrut\\ \hline
JSMA, $\theta$ = 0.01 & \textbf{Fails} & \textbf{Fails} & \textbf{Fails} & \textbf{Fails} & 6349.2156 \TBstrut\\ \hline
\rowcolor{grey}
DeepFool, Default & 17.690 & 13.477 & 255.000 & 0.0020 & 87.4947 \TBstrut\\ \hline
LBFGS, Default & 19.74 & 9.670 & 253.529 & 0.4100 & 4915.2149 \TBstrut\\ \hline
\rowcolor{grey}
BIM, Default & \textbf{Fails} & \textbf{Fails} & \textbf{Fails} & \textbf{Fails} & 450.3401 \TBstrut\\ \hline
PGD, Default & \textbf{Fails} & \textbf{Fails} & \textbf{Fails} & \textbf{Fails} & 6267.9812 \TBstrut\\ \hline
\rowcolor{grey}
C\&W, $c$ = 0 & 19.41 & 9.796 & 239.149 & 0.3200 & 48992.2740 \TBstrut\\ \hline
C\&W, $c$ = 50 & 19.20 & 9.532 & 212.341 & 0.3122 & 47856.1244 \TBstrut\\ \hline
\rowcolor{grey}
C\&W, $c$ = 100 & 19.12 & 9.942 & 239.597 & 0.3000 & 47966.0449 \TBstrut\\ \hline
\end{tabular}
\end{table}
\begin{table}[!h]
\scriptsize
\centering
\caption{EXPERIMENTAL RESULTS FOR ADVERSARY ATTACKS ON A SPRITZ-1.5C (Trained with RIPE Dataset). The time of running attacks on the models is reported in seconds in this table.  \label{SPRITZ_Attack_RIPE}}
\begin{tabular}{|c|c|c|c|c|c|}
\hline
\textbf{Attack Type} & \textbf{PSNR} & \textbf{$L_1$ dist} & \textbf{Max. dist} & \textbf{ASR} & \textbf{Exe. Time} \\ \hline
\rowcolor{grey}
I-FGSM, $\varepsilon$ = 0.1 & \textbf{Fails} & \textbf{Fails} & \textbf{Fails} & \textbf{Fails} & 521.405 \TBstrut\\ \hline
I-FGSM, $\varepsilon$ = 0.01 & \textbf{Fails} & \textbf{Fails} & \textbf{Fails} & \textbf{Fails} & 5304.2130 \TBstrut\\ \hline
\rowcolor{grey}
I-FGSM, $\varepsilon$ = 0.001 & \textbf{Fails} & \textbf{Fails} & \textbf{Fails} & \textbf{Fails} & 66700.5011 \TBstrut\\ \hline
FGSM, $\varepsilon$ = 0.1 & \textbf{Fails} & \textbf{Fails} & \textbf{Fails} & \textbf{Fails} & 69.0314 \TBstrut\\ \hline
\rowcolor{grey}
FGSM, $\varepsilon$ = 0.01 & \textbf{Fails} & \textbf{Fails} & \textbf{Fails} & \textbf{Fails} & 80.9420 \TBstrut\\ \hline
FGSM, $\varepsilon$ = 0.001 & \textbf{Fails} & \textbf{Fails} & \textbf{Fails} & \textbf{Fails} & 372.1473 \TBstrut\\ \hline
\rowcolor{grey}
JSMA, $\theta$ = 0.1 & \textbf{Fails} & \textbf{Fails} & \textbf{Fails} & \textbf{Fails} & 6200.8241 \TBstrut\\ \hline
JSMA, $\theta$ = 0.01 & \textbf{Fails} & \textbf{Fails} & \textbf{Fails} & \textbf{Fails} & 9215.1270 \TBstrut\\ \hline
\rowcolor{grey}
DeepFool, Default & 18.60 & 14.515 & 255.100 & 0.0041 & 91.3950 \TBstrut\\ \hline
LBFGS, Default & 20.43 & 10.310 & 250.412 & 0.2010 & 3420.1860 \TBstrut\\ \hline
\rowcolor{grey}
BIM, Default & 20.00 & 10.421 & 251.326 & 0.0020 & 310.4514 \TBstrut\\ \hline
PGD, Default & 22.12 & 12.67 & 250.120 & 0.1012 & 7130.8523 \TBstrut\\ \hline
\rowcolor{grey}
C\&W, $c$ = 0 & 18.21 & 8.142 & 221.020 & 0.2200 & 48992.2740 \TBstrut\\ \hline
C\&W, $c$ = 50 & 18.10 & 8.411 & 214.190 & 0.2101 & 43871.0109 \TBstrut\\ \hline
\rowcolor{grey}
C\&W, $c$ = 100 & 18.00 & 8.820 & 231.475 & 0.2000 & 48540.1309 \TBstrut\\ \hline
\end{tabular}
\end{table}
\indent The experimental results of the attacks on the SPRITZ-1.5C classifier when trained with an N-BaIoT dataset are shown in Table \ref{SPRITZ_Attack} and in Table \ref{SPRITZ_Attack_RIPE} while trained with a RIPE dataset. In this scenario, the attack usually requires more iteration to enter the pristine region. However, ASR demonstrates that even with high attack iteration, SPRITZ-1.5C is still secure against various adversarial attacks since most of the adversarial attacks completely fail to fool a model. According to the experimental results obtained with the N-BaIoT dataset in Table \ref{SPRITZ_Attack}, all attacks, including I-FGSM, FGSM, JSMA, BIM, and PGD, Fails to mislead a network. Only DeepFool, LBFGS, and CW adversarial attacks can fool the SPRITZ-1.5C; as we indicated, if ASR is less than 50\%, the system is still secure. To elaborate more, in C\&Ws, the ASR is close to 30\%. In only one attack with LBFGS, the ASR is 40\%, which is still below 50\%. As a result, we can state that 1C-Leg and 1C-Mal are quite effective in strengthening the robustness of SPRITZ-1.5C, indicating that the 1C classifiers are more difficult to attack as a consequence of the use of a closed acceptance region. Furthermore, based on the execution running time, attacks,  even with a significant distortion into the examples, Fails to mislead a SPRITZ-1.5C. \\
\indent According to the experimental results obtained with a RIPE dataset, in Table \ref{SPRITZ_Attack_RIPE}, the SPRITZ-1.5C is still robust against various adversarial attacks. All attacks, including I-FGSM, FGSM, and JSMA, completely Fails to deceive a network. In Table \ref{SPRITZ_Attack_RIPE}, for the adversarial attacks such as DeepFool and BIM attacks, the ASR is close to 0\%, implying that the network is fairly secure against these attacks. On the other hand, with LBFGS, PGD, and C\&Ws attacks, ASR is close to 20\%, meaning that only 20\% of attack examples were misclassified, so the ASR is lower than 50\%, suggesting that the system remains secure.


\section{Conclusion and Future Work}
\label{Conclusions}

According to recent studies, most ML and DL methods are intrinsically vulnerable and fragile to adversarial attacks, posing new serious security threats to cybersecurity tools. The study of the security of ML and DL-based methods in the presence of an adversary becomes important. As a result, in this study, we proposed the SPRITZ-1.5C classifier, a multi-classifier architecture, to mitigate the damage caused by an adversary in a PK scenario. The SPRITZ-1.5C classification techniques successfully take the benefits of 2C and 1C techniques, providing better security while maintaining 2C classification's exceptional performance in the absence of attacks. We trained our classifier to discriminate between pristine and malicious examples using the prominent datasets namely: N-BaIoT and RIPE-Atlas.
The experimental results demonstrated the robustness of SPRITZ-1.5C against several adversarial attacks. In our study, we opted for exploratory types of attacks over causative attacks since most of the CF attacks that have been provided currently fit into the category of exploratory attacks. \\
\indent In a future work, we aim to develop SPRITZ-1.5C by employing a block-GAN combination. As a result, it would be interesting to identify the most suitable GAN network for this research study. Another possible future study direction is the development of backdoor or poisoning attacks that adversaries may employ to interrupt the training phase, often known as causative attacks. 

\bibliographystyle{IEEEtran}
\bibliography{References}

\begin{thebibliography}{10}
\providecommand{\url}[1]{#1}
\csname url@samestyle\endcsname
\providecommand{\newblock}{\relax}
\providecommand{\bibinfo}[2]{#2}
\providecommand{\BIBentrySTDinterwordspacing}{\spaceskip=0pt\relax}
\providecommand{\BIBentryALTinterwordstretchfactor}{4}
\providecommand{\BIBentryALTinterwordspacing}{\spaceskip=\fontdimen2\font plus
\BIBentryALTinterwordstretchfactor\fontdimen3\font minus
  \fontdimen4\font\relax}
\providecommand{\BIBforeignlanguage}[2]{{%
\expandafter\ifx\csname l@#1\endcsname\relax
\typeout{** WARNING: IEEEtran.bst: No hyphenation pattern has been}%
\typeout{** loaded for the language `#1'. Using the pattern for}%
\typeout{** the default language instead.}%
\else
\language=\csname l@#1\endcsname
\fi
#2}}
\providecommand{\BIBdecl}{\relax}
\BIBdecl

\bibitem{Gloe2007}
T.~Gloe, M.~Kirchner, A.~Winkler, and R.~B\"{o}hme, ``Can we trust digital
  image forensics?'' in \emph{Proceedings of the 15th ACM International
  Conference on Multimedia}, ser. MM '07.\hskip 1em plus 0.5em minus
  0.4em\relax New York, NY, USA: Association for Computing Machinery, 2007, p.
  78–86.

\bibitem{bohme2013counter}
R.~B{\"o}hme and M.~Kirchner, ``Counter-forensics: Attacking image forensics,''
  in \emph{Digital image forensics}.\hskip 1em plus 0.5em minus 0.4em\relax
  Springer, 2013, pp. 327--366.

\bibitem{nowroozi2020machine}
E.~Nowroozi, M.~Barni, and B.~Tondi, ``Machine learning techniques for image
  forensics in adversarial setting,'' Ph.D. dissertation, 2020.

\bibitem{NOWROOZI2021102092}
E.~Nowroozi, A.~Dehghantanha, R.~M. Parizi, and K.-K.~R. Choo, ``A survey of
  machine learning techniques in adversarial image forensics,'' \emph{Computers
  \& Security}, vol. 100, p. 102092, 2021.

\bibitem{Ehsan_URL_Detection}
\BIBentryALTinterwordspacing
E.~Nowroozi, {Abhishek}, M.~Mohammadi, and M.~Conti, ``An adversarial attack
  analysis on malicious advertisement url detection framework,'' 2022.
  [Online]. Available: \url{https://arxiv.org/abs/2204.13172}
\BIBentrySTDinterwordspacing

\bibitem{hernandez2022adversarial}
C.~J. Hern{\'a}ndez-Castro, Z.~Liu, A.~Serban, I.~Tsingenopoulos, and
  W.~Joosen, ``Adversarial machine learning,'' in \emph{Security and Artificial
  Intelligence}.\hskip 1em plus 0.5em minus 0.4em\relax Springer, 2022, pp.
  287--312.

\bibitem{Ehsan_Eusipco_2017}
M.~Barni, E.~Nowroozi, and B.~Tondi, ``Higher-order, adversary-aware, double
  jpeg-detection via selected training on attacked samples,'' in \emph{2017
  25th European Signal Processing Conference (EUSIPCO)}, 2017, pp. 281--285.

\bibitem{Ehsan_IWBF}
------, ``Detection of adaptive histogram equalization robust against jpeg
  compression,'' in \emph{2018 International Workshop on Biometrics and
  Forensics (IWBF)}, 2018, pp. 1--8.

\bibitem{kang2022using}
S.~Kang, ``Using binary classifiers for one-class classification,''
  \emph{Expert Systems with Applications}, vol. 187, p. 115920, 2022.

\bibitem{biggio2013evasion}
B.~Biggio, I.~Corona, D.~Maiorca, B.~Nelson, N.~{\v{S}}rndi{\'c}, P.~Laskov,
  G.~Giacinto, and F.~Roli, ``Evasion attacks against machine learning at test
  time,'' in \emph{Joint European conference on machine learning and knowledge
  discovery in databases}.\hskip 1em plus 0.5em minus 0.4em\relax Springer,
  2013, pp. 387--402.

\bibitem{barni2020improving}
M.~Barni, E.~Nowroozi, and B.~Tondi, ``Improving the security of image
  manipulation detection through one-and-a-half-class multiple
  classification,'' \emph{Multimedia Tools and Applications}, vol.~79, no.~3,
  pp. 2383--2408, 2020.

\bibitem{analysis}
\BIBentryALTinterwordspacing
``{RIPE-Atlas-Community, "Ripe-Atlas-Community/Ripe-Atlas-data-analysis"}.''
  [Online]. Available:
  \url{https://github.com/RIPE-Atlas-Community/RIPE-Atlas-data-analysis}
\BIBentrySTDinterwordspacing

\bibitem{nbaiot}
\BIBentryALTinterwordspacing
``{UCI Machine Learning Repository:
  detection{\_}of{\_}IoT{\_}botnet{\_}attacks{\_}N{\_}BaIoT Data Set}.''
  [Online]. Available:
  \url{https://archive.ics.uci.edu/ml/datasets/detection_of_IoT_botnet_attacks_N_BaIoT}
\BIBentrySTDinterwordspacing

\bibitem{NBaIoT_Auto}
Y.~Meidan, M.~Bohadana, Y.~Mathov, Y.~Mirsky, A.~Shabtai, D.~Breitenbacher, and
  Y.~Elovici, ``N-baiot—network-based detection of iot botnet attacks using
  deep autoencoders,'' \emph{IEEE Pervasive Computing}, vol.~17, no.~3, pp.
  12--22, 2018.

\bibitem{BaIoT_Anomaly}
F.~Abbasi, M.~Naderan, and S.~E. Alavi, ``Anomaly detection in internet of
  things using feature selection and classification based on logistic
  regression and artificial neural network on n-baiot dataset,'' in \emph{5th
  International Conference on Internet of Things and Applications (IoT)}, 2021,
  pp. 1--7.

\bibitem{REY2022108693}
V.~Rey, P.~M. {Sánchez Sánchez}, A.~{Huertas Celdrán}, and G.~Bovet,
  ``Federated learning for malware detection in iot devices,'' \emph{Computer
  Networks}, vol. 204, p. 108693, 2022.

\bibitem{Turning2017}
\BIBentryALTinterwordspacing
K.~Angrishi, ``Turning internet of things(iot) into internet of vulnerabilities
  (iov) : Iot botnets,'' \emph{CoRR}, vol. abs/1702.03681, 2017. [Online].
  Available: \url{http://arxiv.org/abs/1702.03681}
\BIBentrySTDinterwordspacing

\bibitem{AHANGER2022108771}
T.~A. Ahanger, A.~Aljumah, and M.~Atiquzzaman, ``State-of-the-art survey of
  artificial intelligent techniques for iot security,'' \emph{Computer
  Networks}, vol. 206, p. 108771, 2022.

\bibitem{BASHLITEThreatpost}
\BIBentryALTinterwordspacing
``{BASHLITE Family Of Malware Infects 1 Million IoT Devices | Threatpost}.''
  [Online]. Available:
  \url{https://threatpost.com/bashlite-family-of-malware-infects-1-million-iot-devices/120230/}
\BIBentrySTDinterwordspacing

\bibitem{tahaei2020rise}
H.~Tahaei, F.~Afifi, A.~Asemi, F.~Zaki, and N.~B. Anuar, ``The rise of traffic
  classification in iot networks: A survey,'' \emph{Journal of Network and
  Computer Applications}, vol. 154, p. 102538, 2020.

\bibitem{analysis2}
\BIBentryALTinterwordspacing
``{Secure Machine Learning for Network Analytics}.'' [Online]. Available:
  \url{https://labs.ripe.net/author/andreas_dewes/secure-machine-learning-for-network-analytics/}
\BIBentrySTDinterwordspacing

\bibitem{zhang2019adversarial}
J.~Zhang and C.~Li, ``Adversarial examples: Opportunities and challenges,''
  \emph{IEEE transactions on neural networks and learning systems}, vol.~31,
  no.~7, pp. 2578--2593, 2019.

\bibitem{aldahdooh2022adversarial}
A.~Aldahdooh, W.~Hamidouche, S.~A. Fezza, and O.~D{\'e}forges, ``Adversarial
  example detection for dnn models: a review and experimental comparison,''
  \emph{Artificial Intelligence Review}, pp. 1--60, 2022.

\bibitem{FGSM}
\BIBentryALTinterwordspacing
I.~J. Goodfellow, J.~Shlens, and C.~Szegedy, ``Explaining and harnessing
  adversarial examples,'' 2014. [Online]. Available:
  \url{https://arxiv.org/abs/1412.6572}
\BIBentrySTDinterwordspacing

\bibitem{szegedy2013intriguing}
C.~Szegedy, W.~Zaremba, I.~Sutskever, J.~Bruna, D.~Erhan, I.~Goodfellow, and
  R.~Fergus, ``Intriguing properties of neural networks,'' \emph{arXiv preprint
  arXiv:1312.6199}, 2013.

\bibitem{madry2017towards}
A.~Madry, A.~Makelov, L.~Schmidt, D.~Tsipras, and A.~Vladu, ``Towards deep
  learning models resistant to adversarial attacks,'' \emph{arXiv preprint
  arXiv:1706.06083}, 2017.

\bibitem{kurakin2018adversarial}
A.~Kurakin, I.~J. Goodfellow, and S.~Bengio, ``Adversarial examples in the
  physical world,'' in \emph{Artificial intelligence safety and
  security}.\hskip 1em plus 0.5em minus 0.4em\relax Chapman and Hall/CRC, 2018,
  pp. 99--112.

\bibitem{Moosavi-DezfooliDeepFool:Networks}
\BIBentryALTinterwordspacing
S.~M. Moosavi-Dezfooli, A.~Fawzi, and P.~Frossard, ``{DeepFool: A Simple and
  Accurate Method to Fool Deep Neural Networks},'' \emph{Proceedings of the
  IEEE Computer Society Conference on Computer Vision and Pattern Recognition},
  vol. 2016-Decem, pp. 2574--2582, 2016. [Online]. Available:
  \url{http://github.com/lts4/deepfool}
\BIBentrySTDinterwordspacing

\bibitem{deepfoolcite}
Y.~Ding, G.~Zhu, D.~Chen, X.~Qin, M.~Cao, and Z.~Qin, ``Adversarial sample
  attack and defense method for encrypted traffic data,'' \emph{IEEE
  Transactions on Intelligent Transportation Systems}, pp. 1--16, 2022.

\bibitem{carlini2017towards}
N.~Carlini and D.~Wagner, ``Towards evaluating the robustness of neural
  networks,'' in \emph{2017 ieee symposium on security and privacy (sp)}.\hskip
  1em plus 0.5em minus 0.4em\relax IEEE, 2017, pp. 39--57.

\bibitem{d2017autoencoder}
D.~D'Avino, D.~Cozzolino, G.~Poggi, and L.~Verdoliva, ``Autoencoder with
  recurrent neural networks for video forgery detection,'' \emph{Electronic
  Imaging}, vol. 2017, no.~7, pp. 92--99, 2017.

\bibitem{marchi2015novel}
E.~Marchi, F.~Vesperini, F.~Eyben, S.~Squartini, and B.~Schuller, ``A novel
  approach for automatic acoustic novelty detection using a denoising
  autoencoder with bidirectional lstm neural networks,'' in \emph{2015 IEEE
  international conference on acoustics, speech and signal processing
  (ICASSP)}.\hskip 1em plus 0.5em minus 0.4em\relax IEEE, 2015, pp. 1996--2000.

\bibitem{xu2022improved}
W.~Xu, J.~Jang-Jaccard, T.~Liu, F.~Sabrina, and J.~Kwak, ``Improved
  bidirectional gan-based approach for network intrusion detection using
  one-class classifier,'' \emph{Computers}, vol.~11, no.~6, p.~85, 2022.

\bibitem{perdisci2006using}
R.~Perdisci, G.~Gu, and W.~Lee, ``Using an ensemble of one-class svm
  classifiers to harden payload-based anomaly detection systems,'' in
  \emph{Sixth International Conference on Data Mining (ICDM'06)}.\hskip 1em
  plus 0.5em minus 0.4em\relax IEEE, 2006, pp. 488--498.

\bibitem{scheirer2012toward}
W.~J. Scheirer, A.~de~Rezende~Rocha, A.~Sapkota, and T.~E. Boult, ``Toward open
  set recognition,'' \emph{IEEE transactions on pattern analysis and machine
  intelligence}, vol.~35, no.~7, pp. 1757--1772, 2012.

\bibitem{huang2022novel}
K.~Huang, J.~Yang, P.~Hu, and H.~Liu, ``A novel framework for open-set
  authentication of internet of things using limited devices,'' \emph{Sensors},
  vol.~22, no.~7, p. 2662, 2022.

\bibitem{henrydoss2017incremental}
J.~Henrydoss, S.~Cruz, E.~M. Rudd, M.~Gunther, and T.~E. Boult, ``Incremental
  open set intrusion recognition using extreme value machine,'' in \emph{2017
  16th IEEE International Conference on Machine Learning and Applications
  (ICMLA)}.\hskip 1em plus 0.5em minus 0.4em\relax IEEE, 2017, pp. 1089--1093.

\bibitem{sawadogo2022android}
Z.~Sawadogo, G.~Mendy, J.~M. Dembelle, and S.~Ouya, ``Android malware
  classification: Updating features through incremental learning approach
  (ufila),'' in \emph{2022 24th International Conference on Advanced
  Communication Technology (ICACT)}.\hskip 1em plus 0.5em minus 0.4em\relax
  IEEE, 2022, pp. 544--550.

\bibitem{akgun2022new}
D.~AKGUN, S.~HIZAL, and U.~CAVUSOGLU, ``A new ddos attacks intrusion detection
  model based on deep learning for cybersecurity,'' \emph{Computers \&
  Security}, p. 102748, 2022.

\bibitem{gumusbas2022ai}
D.~Gumusbas and T.~Yildirim, ``Ai for cybersecurity: Ml-based techniques for
  intrusion detection systems,'' in \emph{Advances in Machine Learning/Deep
  Learning-based Technologies}.\hskip 1em plus 0.5em minus 0.4em\relax
  Springer, 2022, pp. 117--140.

\bibitem{MOTI2021102591}
Z.~Moti, S.~Hashemi, H.~Karimipour, A.~Dehghantanha, A.~N. Jahromi, L.~Abdi,
  and F.~Alavi, ``Generative adversarial network to detect unseen internet of
  things malware,'' \emph{Ad Hoc Networks}, vol. 122, p. 102591, 2021.

\bibitem{9747933}
E.~Nowroozi, Y.~Mekdad, M.~H. Berenjestanaki, M.~Conti, and A.~EL~Fergougui,
  ``Demystifying the transferability of adversarial attacks in computer
  networks,'' \emph{IEEE Transactions on Network and Service Management}, pp.
  1--1, 2022.

\bibitem{9053318}
M.~Barni, E.~Nowroozi, B.~Tondi, and B.~Zhang, ``Effectiveness of random deep
  feature selection for securing image manipulation detectors against
  adversarial examples,'' in \emph{ICASSP 2020 - 2020 IEEE International
  Conference on Acoustics, Speech and Signal Processing (ICASSP)}, 2020, pp.
  2977--2981.

\bibitem{dasgupta2022machine}
D.~Dasgupta, Z.~Akhtar, and S.~Sen, ``Machine learning in cybersecurity: a
  comprehensive survey,'' \emph{The Journal of Defense Modeling and
  Simulation}, vol.~19, no.~1, pp. 57--106, 2022.

\bibitem{8451698}
M.~Barni, A.~Costanzo, E.~Nowroozi, and B.~Tondi, ``Cnn-based detection of
  generic contrast adjustment with jpeg post-processing,'' in \emph{2018 25th
  IEEE International Conference on Image Processing (ICIP)}, 2018, pp.
  3803--3807.

\bibitem{SPRITZ15C}
E.~Nowroozi, ``Spritz-1.5c,''
  \url{https://github.com/ehsannowroozi/SPRITZ-1.5C/blob/main/README.md}, 2022.

\end{thebibliography}

\vskip -3\baselineskip plus -1fil
\begin{IEEEbiography}
[{\includegraphics[width=1in,height=1.25in]{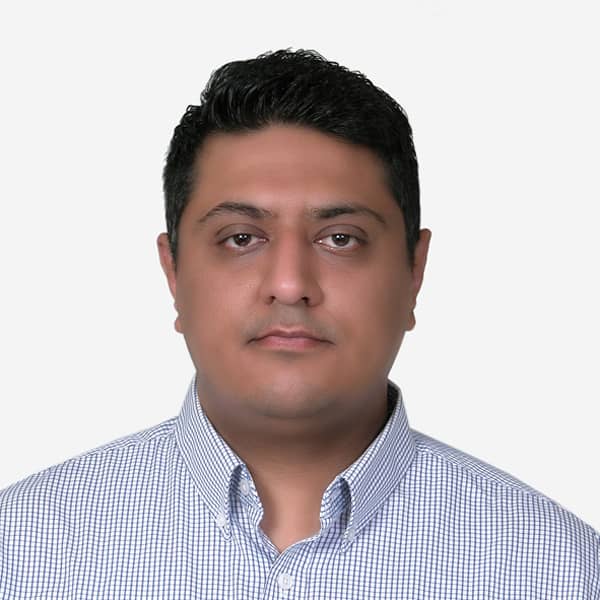}}]
{Ehsan Nowroozi} is an Assistant Professor at Istanbul’s Bahcesehir University’s (BAU) Faculty of Engineering and Natural Sciences, Department of Computer Engineering. In 2020, he received his Ph.D. from the University of Siena in Italy Following his Ph.D., he was a postdoctoral fellow at Siena and Padua Universities in Italy and Sabanci University in Turkey. His primary research interests are in Cybersecurity in a particular reference to adversarial machine learning and adversarial multimedia forensics. He is also a reviewer for several journals, including IEEE TNSM, IEEE TIFS, IEEE TNNLS, and so on. He is a Senior member of the IEEE since 2022 till now. \\ 
\end{IEEEbiography}

\vskip -2\baselineskip plus -1fil
\begin{IEEEbiography}[{\includegraphics[width=1in,height=1.25in]{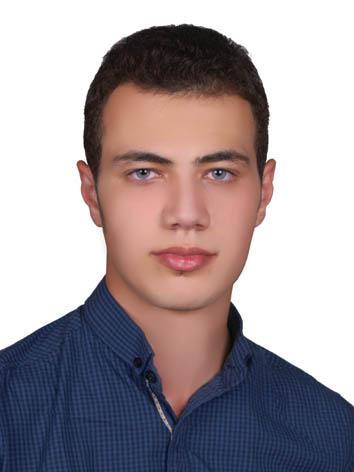}}]{Mohammadreza Mohammadi} is a second-year Master student of ICT at University of Padua. He received the bachelor’s degree in computer engineering(software-network) from Bu-Ali Sina University, Hamedan, Iran in 2019. His main research interest is in the area of Machine Learning, Cybersecurity, IoT and Computer Vision. In addition, he works on reserach contexts which are combination of Industrial IoT security and artificial intelligence, and Intrusion detection systems(IDS) and Healthcare systems and He is also a graduate student member of IEEE institution.
\end{IEEEbiography}

\vskip -2\baselineskip plus -1fil
\begin{IEEEbiography}[{\includegraphics[width=1in,height=1.25in]{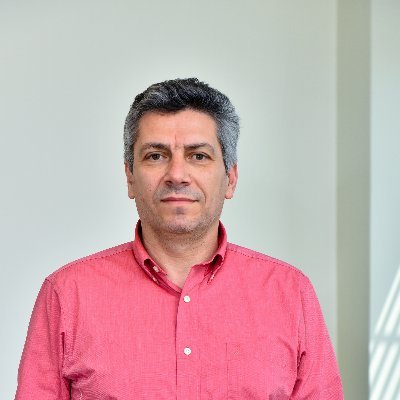}}]{Erkay Savaş} received the BS and MS degrees in electrical engineering from the Electronics and Communications Engineering Department, Istanbul Technical University in 1990 and 1994, respectively. He received the PhD degree from the
Department of Electrical and Computer Engineering, Oregon
State University in June 2000. He has been a faculty member at Sabanci University since 2002. His research interests include applied cryptography, data and communication security, security and privacy in data mining applications, embedded systems security, and distributed systems. He is a member of IEEE, ACM, the IEEE Computer Society, and the International Association of Cryptologic Research. K. Derya et al.
\end{IEEEbiography}

\vskip -2\baselineskip plus -1fil
\begin{IEEEbiography}[{\includegraphics[width=1in,height=1.25in]{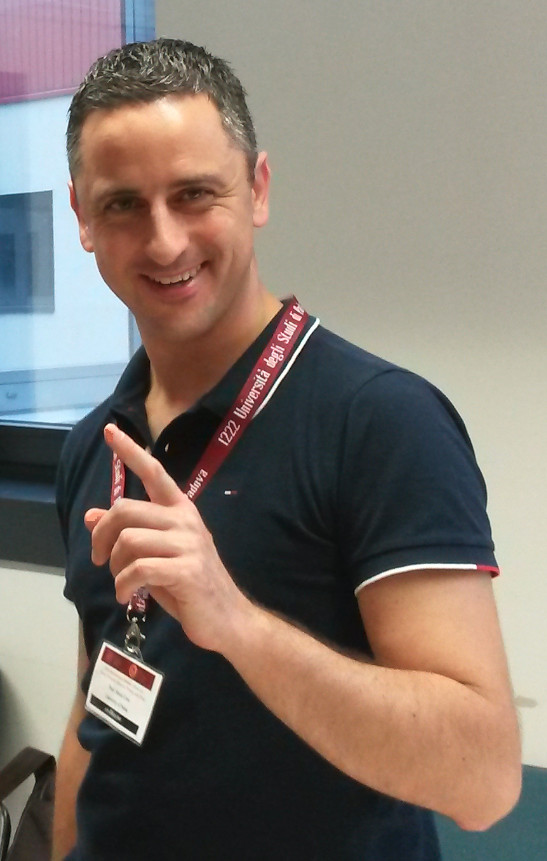}}]{Mauro Conti} is Full Professor at the University of Padua, Italy. He is also affiliated with TU Delft and University of Washington, Seattle. He obtained his Ph.D. from Sapienza University of Rome, Italy, in 2009. His research in the area of Security and Privacy is also funded by companies, including Cisco, Intel, and Huawei. he published more than 450 papers in topmost international peer-reviewed journals and conferences. He is Editor-in-Chief for IEEE Transactions on Information Forensics and Security, and has been Associate Editor for several journals, including IEEE Communications Surveys \& Tutorials, IEEE Transactions on Dependable and Secure Computing, IEEE Transactions on Network and Service Management. He is Fellow of the IEEE and YAE and Senior member of ACM.
\end{IEEEbiography}

\vskip -2\baselineskip plus -1fil
\begin{IEEEbiography}[{\includegraphics[width=1in,height=1.25in]{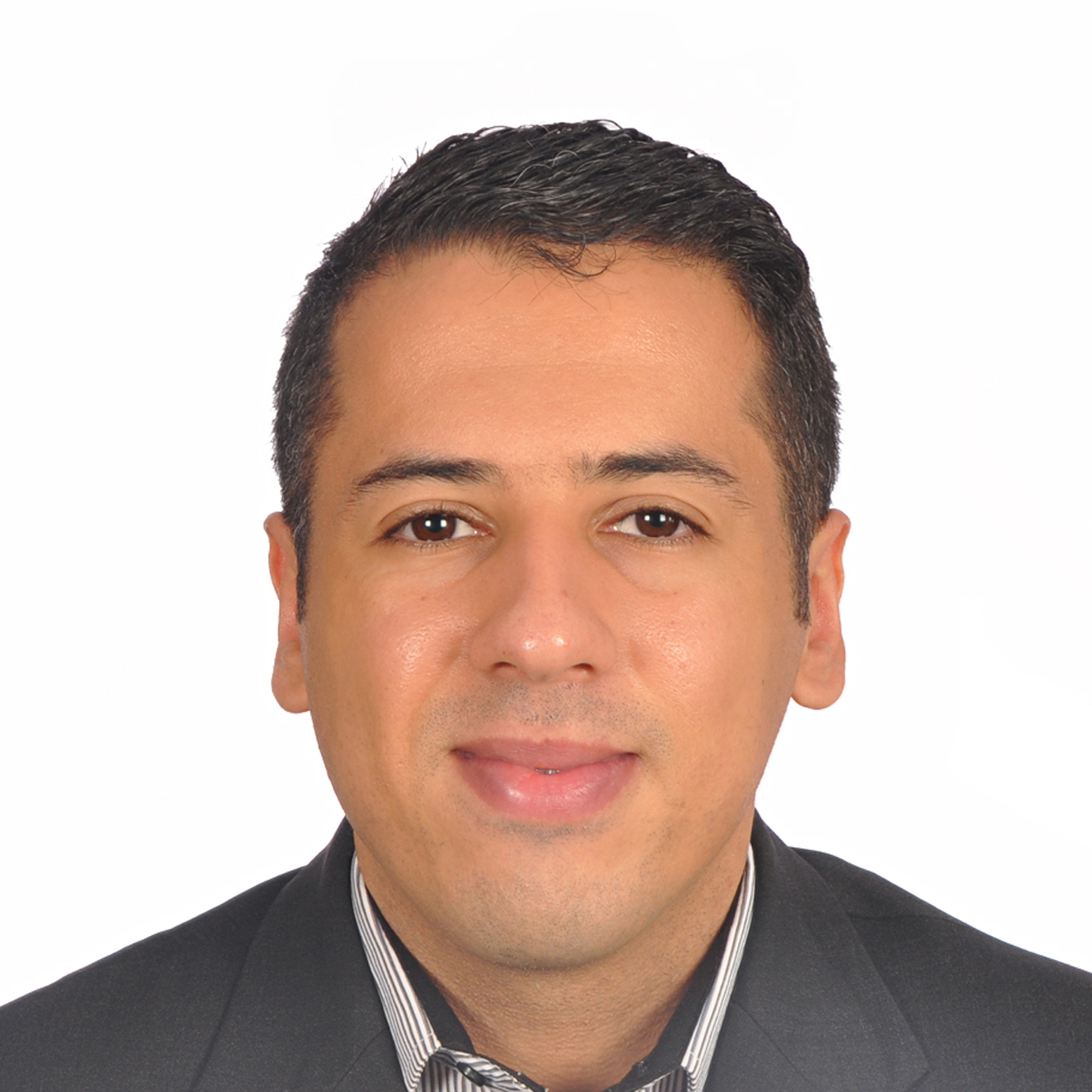}}]{Yassine Mekdad} received a Msc Degree in Cryptography and Information Security from Mohammed V University of Rabat, Morocco. He holds a guest researcher position with the SPRITZ research group at University of Padua, Italy. He is currently working as a Research Scholar at the Cyber-Physical Systems Security Lab (CSL) at Florida International University, Miami, FL, USA. His research interest principally cover security and privacy problems in the Internet of Things (IoT), Industrial Internet-of-Things (IIoT), and Cyber-physical systems (CPS). Furthermore, he works on research problems at the intersection of the cybersecurity and networking fields with an emphasis on their practical and applied aspects. 
\end{IEEEbiography}

\end{document}